\documentclass[11pt]{iopart}
\pdfoutput=1
\usepackage{xcolor}
\eqnobysec
\expandafter\let\csname equation*\endcsname=\relax
\expandafter\let\csname endequation*\endcsname=\relax

\usepackage{amsmath}
\usepackage{enumerate}
\usepackage{graphicx}
\usepackage{mathbbol}
\usepackage{amsfonts}
\usepackage{hyperref}
\newcommand{\lmax}{\ell_{\max}}
\newcommand{\beq}{\begin{eqnarray}}
\newcommand{\eeq}{\end{eqnarray}}
\newcommand{\be}{\begin{equation}}
\newcommand{\ee}{\end{equation}}

\newcommand{\beqa}{\begin{eqnarray}}

\newcommand{\eeqa}{\end{eqnarray}}

\renewcommand{\d}{{\rm d }}

\renewcommand{\max}{{\rm max}}
\newcommand{\prob}{\mathop{\rm Prob}\nolimits}

\renewcommand{\l}{{\ell}}
\newcommand{\B}{\rm (0)}

\newcommand{\erfc}{\mathrm{erfc}}

\def\XXint#1#2#3{{\setbox0=\hbox{$#1{#2#3}{\int}$}
 \vcenter{\hbox{$#2#3$}}\kern-.5\wd0}}

\begin{document}
\title[Record statistics for random walk bridges]{Record statistics for random walk bridges}%Title of paper

\author{Claude Godr\`eche}
\address{Institut de Physique Th\'eorique, Universit\'e Paris-Saclay, CEA and CNRS, 91191 Gif-sur-Yvette, France}

\author{Satya N. Majumdar}
\address{Universit\'e Paris-Sud, LPTMS, CNRS (UMR 8626), 91405 Orsay Cedex, France}

\author{Gr\'egory Schehr}
\address{Universit\'e Paris-Sud, LPTMS, CNRS (UMR 8626), 91405 Orsay Cedex, France}

\begin{abstract}
We investigate the statistics of records in a random sequence 
$\{x_B(0)=0,x_B(1),\cdots, x_B(n)=x_B(0)=0\}$ of $n$ time steps. The 
sequence $x_B(k)$'s represents the position at step $k$ of a random walk 
`bridge' of $n$ steps that starts and ends at the origin. At each step, 
the increment of the position is a random jump drawn from a specified 
symmetric distribution. We study the statistics of records and record ages 
for such a bridge sequence, for different jump distributions. In absence 
of the bridge condition, i.e., for a free random walk sequence, the 
statistics of the number and ages of records exhibits a `strong' 
universality for all $n$, i.e., they are completely independent of the 
jump distribution as long as the distribution is continuous. We show that 
the presence of the bridge constraint destroys this strong `all $n$' 
universality.  Nevertheless a `weaker' universality still remains for 
large $n$, where we show that the record statistics depends on the jump 
distributions only through a {\em single} parameter $0<\mu\le 2$, known as 
the L\'evy index of the walk, but are insensitive to the other details of 
the jump distribution. We derive the most general results (for arbitrary 
jump distributions) wherever possible and also present two exactly 
solvable cases. We present numerical simulations that verify our 
analytical results.
\end{abstract}

\maketitle

\section{Introduction and summary of main results}

During the last few years, there has been a growing interest in the study of records. Records have not only become 
popular in our societies as one often hears and reads, in the media, about record breaking events, including for instance sports \cite{Gembris} or weather records \cite{weather_records_wiki}, but they have also found interesting applications in various areas of sciences. Records are often useful to characterize
the statistics of a (discrete) time series $x(0), x(1), \ldots, x(n)$ where a record happens at time $k$ if $x(k)$ is larger than the previous values $x(0), x(1), \cdots, x(k-1)$. As such, they have been studied in several contexts ranging 
from domain wall dynamics \cite{ABBM}, spin-glasses \cite{Sibani} and random walks \cite{MZ2008,satya_leuven,sanjib,WMS2012,MSW2012,GMS2014} to avalanches \cite{LDW09}, models of stock prices \cite{WMS2012,WBK2011,Cha2015}, growing networks models \cite{GL2008} or the study of global warming \cite{RP2006,WK2010,AK2011} and also in evolutionary biology \cite{krugjain,franke} (see Ref. \cite{gregor_review} for a recent review). 

When considering record statistics, the most natural questions are the following: (a) what is the number of records after time $n$? (b) what is the probability that a record is broken at step $n$? how long does a record survive and in particular what is the age of the longest lasting record? These questions are fully understood in the case where the variables $x(i)$'s are independent and identically distributed (i.i.d.) \cite{FS1954,Nevzorov} (see also Ref. \cite{SM_review} for a short review), although more refined questions about 
first passage properties of records in this i.i.d. case were only recently investigated \cite{BNK13,MBN13}. However, many applications of records in statistical physics have highlighted the importance of strong correlations between the random variables $x(i)$'s, for which much less is known. With that perspective, it was demonstrated rather recently that one-dimensional random walks offer a very useful instance of strongly correlated variables where the impact of correlations on records can be studied analytically \cite{MZ2008,satya_leuven,sanjib,WMS2012,MSW2012,GMS2014}. As we recall it below, a remarkable feature of the record statistics of RWs with continuous jumps is that it is completely universal, i.e., independent of the jump distributions, even for a finite number of steps \cite{MZ2008}. It is thus natural to ask whether this universality still holds for constrained random walks. This question is the main motivation for the present work where we consider random walk bridges, where the walker is constrained to start and end at the same point, which are relevant, for instance, to study periodic correlated time series.

\begin{figure}
\includegraphics[width=\linewidth]{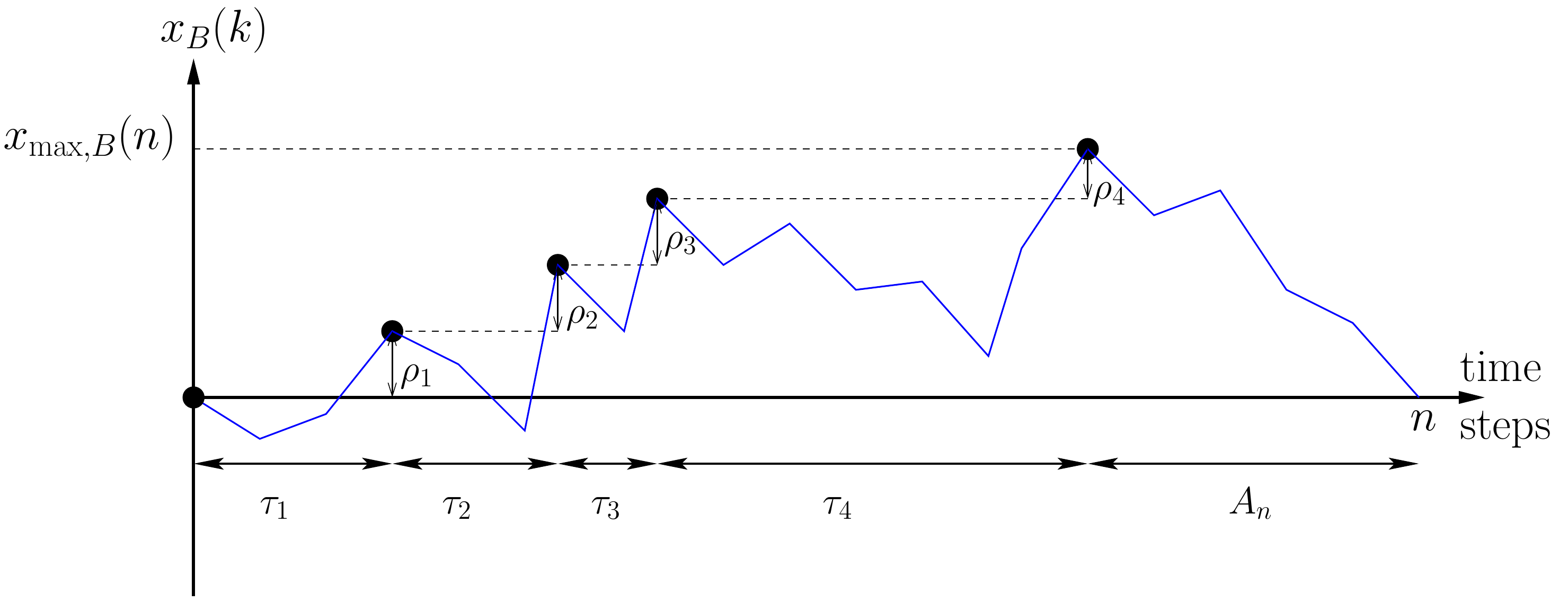}
\caption{A realization of a random walk bridge $x_B(k)$ with an exponential jump distribution ($\alpha = {\rm e}$) with $n=20$ steps. Here the number of records is $R_B^{\rm e}(n) = 5$. The $\tau_i$'s denote the ages of the records and the $\rho_i$'s are the increments between successive records. Finally, $A_n$ denotes the age of the last record. The joint PDF of the $\tau_i$'s, $\rho_i$'s, $A_n$ and $R_B^{\rm e}(n)$ is given in Eq. (\ref{eq:def_joint_pdf_exp_2}) and constitutes one of the main result of this paper.}\label{Fig1}
\end{figure}

Let us consider a time series where the $x(k)$'s are the positions of a random walker (RW) after step $k$, evolving according to the Markov rule:
\begin{eqnarray}\label{def_RW}
x(0) = 0 \;, \; x(k) = x(k-1) + \eta(k) \;, 
\end{eqnarray} 
where $\eta(k)$'s are i.i.d. random variables. We denote by $p(\eta)$ their symmetric probability distribution function (PDF) [with $p(\eta) = p(-\eta)$]. In the following, we will focus on the positions of RW bridges $\{x_B(k)\}$, $0 \leq k \leq n$ which are RWs as defined in (\ref{def_RW}) with the additional constraint that they return back to the origin after $n$ time steps, $x_B(n) = x_B(0) = 0$ (see Fig. \ref{Fig1}). 

For the purpose of our study, it is important to distinguish between jump 
distributions $p(\eta)$ that are respectively discrete and continuous.
A representative of the discrete class is the so called lattice RW
\begin{equation}\label{def_discrete}
p^{\rm d}(\eta)= \frac{1}{2}\, \delta(\eta+1)+ \frac{1}{2}\, 
\delta(\eta-1) \;,
\end{equation}
where the superscript $d$ refers to this particular discrete case. In this 
case, the 
walker moves on a lattice of the integers. In contrast, when the jump 
distribution is continuous, we will denote it generally by a superscript 
$c$, i.e., $p^{\rm c}(\eta)$. 
Let 
$\hat p^{\rm c}(k)= \int_{-\infty}^{\infty} \d\eta\, p^{\rm c}(\eta)\, \e^{i\,k\,\eta}$ 
denote the Fourier transform
of the jump distribution. 
It turns out that for many properties associated with the record 
statistics, only the small $k$-behavior of the jump distribution
matters. Generically, for small $k$, one typically has
\begin{equation}\label{def_p_c}
{\hat p}^{\rm c}(k) = 1- |a\, k|^{\mu} + o(|k|^{\mu}) \;,
\end{equation}
where $0<\mu\le 2$ is called the L\'evy index of the walk and $a>0$ just 
sets the length scale of the jump. For $\mu=2$, the variance 
$\sigma^2=\langle \eta^2\rangle$ is finite. Consequently, $a^2=\sigma^2/2$ 
and the RW converges, for a large number of steps $n$, to the Brownian 
motion. In contrast, for $\mu<2$, the jump distribution has a heavy tail, 
$p^{\rm c}(\eta)\sim |\eta|^{-1-\mu}$ for large $\eta$, leading to
a diverging second moment. A random walk with $\mu<2$ is called a   
L\'evy flight with index $\mu$. Among the class of continuous jump 
distributions with a finite $\sigma^2$, a special role is played 
by the exponential distribution
\begin{equation}\label{def_exp}
p^{\rm e}(\eta)= \frac{1}{2\, b}\, \e^{-|\eta|/b}
\end{equation}
where $b>0$ sets the length scale of the jumps and the superscript `e'
refers to exponential. This clearly belongs to the class in Eq. 
(\ref{def_p_c}) with $\mu=2$ and $a=b$. We will see later that this
exponential distribution belonging to the continuous family,
and the lattice RW in Eq. (\ref{def_discrete}) belonging to the
discrete family, represent two rare cases where the record statistics for
the bridge sequence is exactly solvable.

When studying the sequence of records of a time series $x(0), x(1), \cdots, x(n)$, a record happens at step $k$ if it is {\it strictly} larger than the previous entries, i.e.,
\be\label{def_record}
x(k) > \max(x(0), x(1), \cdots, x(k-1))
\ee
 and by convention, $x(0)$ is a record. We emphasize that the strict inequality in Eq. (\ref{def_record}) is of special importance in the case of discrete random walks (\ref{def_discrete}). We denote by $R(n) \geq 1$ the number of records after $n$ steps. Other important observables for our study are the ages of these records (see Fig.~\ref{Fig1}). We define $\tau_k$ as the number of steps between the $k$-th and $(k+1)$-th records: this is the age of the $k$-th record, i.e. the time up to which the $k$-th record survives. Note that the last record occurring before $n$ is still a record at step $n$ and we denote by $A_n$ the current age of this record at step $n$ (see Fig. \ref{Fig1}). The typical fluctuations of these sequences of ages is rather simple to study. For instance, the typical age of a record is simply related to the average number of records $\langle R(n)\rangle$ {via} $\ell_{\rm typ}\sim n/\langle R(n)\rangle $. However, it has been shown that the statistics of these sequences is dominated, for RWs, by rare events \cite{MZ2008,GMS2014}. To characterize the fluctuations of these rare events, it is useful to consider the probability $Q(n)$ that the last time interval $A_n$ is the longest one, which we call in the following the probability of record breaking. For a RW with a given number of records $R(n) = m$, the sequence of the ages of the records is $\tau_1, \ldots, \tau_{m-1}$ and $A_n$. It is then useful to introduce the joint probability $Q(m,n)$ defined as 
 \beq\label{eq:def_Q1}
 Q(m,n) = {\Pr} (A_n \geq \max(\tau_1, \ldots, \tau_{m-1}), R(n) = m) \;.
 \eeq
 The probability of record breaking $Q(n)$ is then obtained by summing $Q(m,n)$ over all possible number of records in the time series, i.e.,
  \beq\label{eq:def_Q}
 Q(n) = \sum_{m=1}^\infty Q(m,n) \;.
 \eeq
Another important observable to characterize the rare and extreme fluctuations of the sequences of ages is the age of the longest lasting record $\lmax(n)$ which is defined as 
\beq\label{eq:def_lmax}
\lmax(n) = \max(\tau_1, \tau_2, \ldots, \tau_{m-1}, A_n) \;,
\eeq
whose fluctuations, for RWs, were recently demonstrated to be very sensitive to the last record \cite{GMS2014}. 

{\bf Notations:} In the following, we will denote by $R^\alpha(n)$, $Q^\alpha(n)$ and $\ell_{\max}^\alpha(n)$ the quantity defined above for free random walks (\ref{def_RW}) where the superscript $\alpha$ refers to different jump distributions (\ref{def_discrete}, \ref{def_p_c}, \ref{def_exp}), where 
\beq
\hspace*{-1cm}\alpha = 
\begin{cases}
&{\rmd} \;\; {\rm corresponding \; to \; discrete \; jump \; distribution \;} p^{\rmd}(\eta) \;, \\
&{\rm c} \; \; {\rm corresponding \; to \; generic \; continuous \; jump \; distribution \;} p^{\rm c}(\eta) \;, \\
&{\e} \;\; {\rm corresponding \; to \; exponential \; jump \; distribution \;} p^{\e}(\eta) \;, \\
\end{cases}
\eeq 
and where $p^{\rmd}(\eta)$, $p^{\rm c}(\eta)$ and $p^{\e}(\eta)$ are defined respectively in Eq. (\ref{def_discrete}), (\ref{def_p_c}) and (\ref{def_exp}). Besides we will use the notations $R_B^\alpha(n)$, $Q_B^\alpha(n)$ and $\ell_{\max,B}^\alpha(n)$ for the number of records, the probability of record breaking~(\ref{eq:def_Q}) and the age of the longest lasting record (\ref{eq:def_lmax}) for random bridges, where the subscript $B$ refers to bridges.

Before turning to our results for such constrained RWs, let us first remind the main known results for the record statistics of free RWs, where the walker can end up at any position $x_n$ after $n$ time steps~\cite{MZ2008,satya_leuven,sanjib,WMS2012,MSW2012,GMS2014}. We first focus on the case of continuous (symmetric) jump distributions $p^{\rm c}(\eta)$ (\ref{def_p_c}). In this case, a remarkable property is that the full statistics of records, including the number of records $R^{\rm c}(n)$ as well as $Q^{\rm c}(n)$ (\ref{eq:def_Q}) and $\lmax^{\rm c}(n)$ (\ref{eq:def_lmax}) are completely universal, i.e., independent of the jump distribution $p^{\rm c}(\eta)$ -- including L\'evy flights -- even for finite $n$ \cite{MZ2008}. This universal behavior stems from the universality of the Sparre Andersen theorem \cite{SA53}. 
In particular, for large $n$, the average number of records $\langle R^{\rm c}(n)\rangle$ behaves for large $n$ as \cite{MZ2008}
\beq\label{eq:av_records_free}
\langle R^{\rm c}(n)\rangle \sim A^{\rm c} \sqrt{n} \;, \; A^{\rm c} = \frac{2}{\sqrt{\pi}} \;, \; {\rm as} \; \; n \to \infty \;,
\eeq
independently of $0 < \mu \leq 2$ characterizing the jump distribution $p^{\rm c}(\eta)$ in (\ref{def_p_c}). This behavior (\ref{eq:av_records_free}) should be compared to the logarithmic growth found for i.i.d. random variables~\cite{Nevzorov}. Moreover, it is also possible to compute exactly the full probability distribution $P^{\rm c}(m,n) = \Pr(R^{\rm c}(n)=m,n)$ of the number of records which, for large $n$ and large $m$ keeping $X = m/\sqrt{n}$ fixed, takes the scaling form \cite{MZ2008}
\beq\label{eq:distrib_R_free}
P^{\rm c}(m,n) \sim \frac{1}{\sqrt{n}} \varphi^{\rm c}\left(X = \frac{m}{\sqrt{n}} \right) \; {\rm where} \; \varphi^{\rm c} \left( X\right) = \frac{1}{\sqrt{\pi}} \e^{-\frac{X^2}{4}}\;, X > 0 \;.
\eeq
For free random walks, it is also possible to compute exactly $Q^{\rm c}(n)$ in Eq. (\ref{eq:def_Q}) and show that it converges, for large $n$, to a non-trivial universal constant given by \cite{MZ2008}
\beq\label{eq:qinf_free}
\lim_{n \to \infty} Q^{\rm c}(n) = Q^{\rm c}(\infty) = \int_0^\infty \rmd x \frac{1}{1+ \sqrt{\pi \,x} \, \e^x \, {\rm erf}(\sqrt{x})} = 0.626508\ldots,
\eeq
a constant which also arises in the excursion theory of Brownian motion \cite{GMS2009,PY97}. Interestingly, in the case of free RWs, this probability $Q^{\rm c}(n)$ turns out to be related to the average value of $\langle \ell^{\rm c}_{\max}(n) \rangle$ in Eq. (\ref{eq:def_lmax}) via the relation \cite{GMS2009}
\beq\label{eq:relQ_lmax}
\langle \ell^{\rm c}_{\max}(n+1)\rangle = \langle \ell^{\rm c}_{\max}(n)\rangle + Q^{\rm c}(n) \;, 
\eeq
implying in particular that
\beq\label{eq:asympt_lmax}
\hspace*{-0cm}\lim_{n \to \infty} \frac{\langle \ell^{\rm c}_{\max}(n)\rangle}{n} = \lambda_{\max}^{\rm c} = Q^{\rm c}(\infty)=0.626508\ldots \;,
\eeq
while the full distribution of $\ell^{\rm c}_{\max}(n)/n$ was studied more recently in \cite{GMS2015} (see also \cite{Lamp61}). 

\begin{table}
\centering
\includegraphics[width=0.8\linewidth]{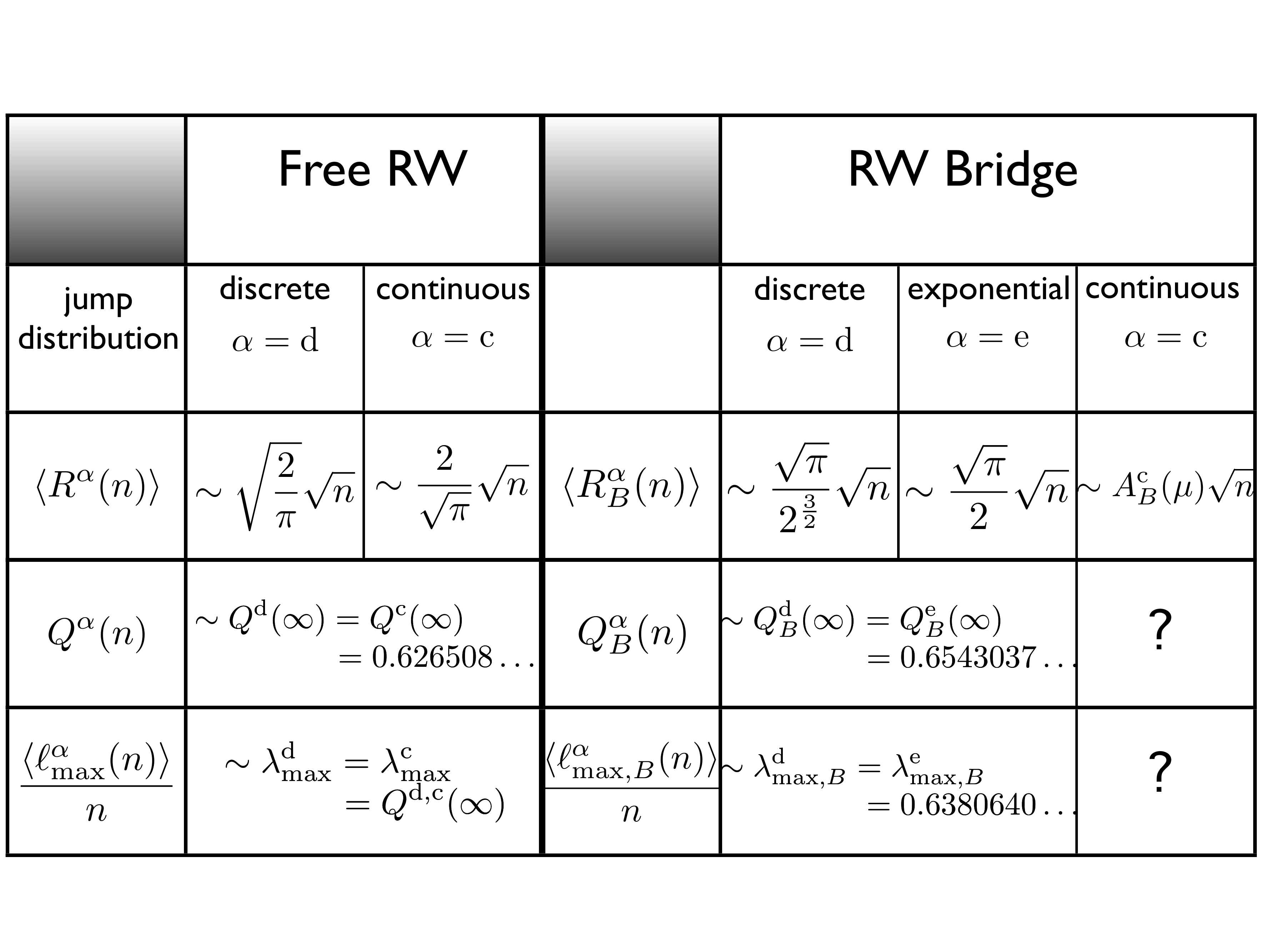}
\caption{Summary of the results for the record statistics of RW bridges, obtained in the present paper. For comparison, we have also presented, in the left part of the table, the results for free RW, obtained in Refs. \cite{MZ2008}. Note that, for the free RW, the results for continuous distributions are completely universal, and thus also hold for an exponential distribution, which is not true for the RW bridges. The expression of the constants $Q_B^{\rm d,e}(\infty)$ and $\lambda_{\max, B}^{\rm d,e}$ is given in Eqs. (\ref{eq:cQ}) and (\ref{eq:const_lmax_discrete}) respectively. The computation of $Q_B^{\rm c}(n)$ and of $\langle \ell_{\max,B}^{\rm c}(n) \rangle$ for a continuous distribution with arbitrary L\'evy index $\mu$ [see Eq. (\ref{def_p_c})]  remains an open question -- hence the question marks (?) in the table.}
\label{table:summary}
\end{table}

For discrete free random walks, with a jump distribution given in Eq. (\ref{def_discrete}), it was shown in \cite{MZ2008} that the statistics of
records is quantitatively different. In particular, in this case, the mean number of records $\langle R^{\rmd}(n) \rangle$ also grows like $\sqrt{n}$ but with a different prefactor
\beq\label{eq:av_records_free_dicrete}
\langle R^{\rmd}(n) \rangle \sim A^{\rmd} \sqrt{n} \;, \; A^{\rmd} = \sqrt\frac{2}{\pi} \;, \; {\rm as \;} n \to \infty \:,
\eeq
which is $1/\sqrt{2}$ of the expression for the mean in the continuous case (\ref{eq:av_records_free}). Similarly, the distribution of $R^{\rmd}(n)$, $P^{\rmd}(m,n)
= \Pr(R^{\rmd}(n)=m,n)$, takes the following scaling form, for large $n$ and large $m$ keeping $X = m/\sqrt{n}$ fixed \cite{MZ2008,WMS2012}:
\beq\label{eq:distrib_R_free_discrete}
P^{\rmd}(m,n) \sim \frac{1}{\sqrt{n}} \varphi^{\rm d}\left(X = \frac{m}{\sqrt{n}} \right) \;, \; \varphi^{\rm d} \left( X\right) = \sqrt{\frac{2}{\pi}} \e^{-\frac{X^2}{2}}\;, \; X > 0 \;,
\eeq
which is simply obtained from its continuous counterpart $\varphi^{\rm c}(X)$ in Eq. (\ref{eq:distrib_R_free}) with the substitution $X \to \sqrt{2}\, X$. On the other hand, the results for the probability of record breaking $Q^{\rmd}(n)$ behaves, for large $n$, exactly as in the continuous case (\ref{eq:qinf_free}), i.e.
\beq\label{eq:qinf_free_discrete}
\hspace*{-1.8cm}\lim_{n \to \infty} Q^{\rmd}(n) = Q^{\rmd}(\infty) = Q^{\rm c}(\infty) = \int_0^\infty \rmd x \frac{1}{1+ \sqrt{\pi \,x} \, \e^x \, {\rm erf}(\sqrt{x})} = 0.626508\ldots \;.
\eeq 
Moreover, the relation in Eq. (\ref{eq:relQ_lmax}) also holds for the corresponding discrete quantities, from which it follows, using Eq. (\ref{eq:qinf_free_discrete}), that
\beq\label{eq:asympt_lmax_discrete}
\hspace*{-0cm}\lim_{n \to \infty} \frac{\langle \ell^{\rm d}_{\max}(n)\rangle}{n} = \lambda^{\rmd}_{\max} = Q^{\rm d}(\infty)= 0.626508\ldots \;.
\eeq

For random walk bridges, the situation is quite different. As expected, the statistics of records for discrete (\ref{def_discrete}) and continuous (\ref{def_p_c}) jump distributions are still different, as they are for free random walks. But in this case, for continuous distributions, the statistics of records, for finite $n$, is not universal any more and depends on $p^{\rm c}(\eta)$. One expects however, and we can show it explicitly for the average number of records $\langle R_B^{\rm c}(n)\rangle$, that, for large $n$,
the various observables characterizing the record statistics depend only on the exponent $\mu$ (\ref{def_p_c}) and not on the further microscopic details of the jump distribution $p^{\rm c}(\eta)$. Another important feature of the record statistics of random walk bridges is that it is technically much more involved. Indeed, for free random walks, the computations require the full joint distribution of the ages of the records $\tau_1, \tau_2, \cdots, \tau_{m-1}, A_n$ but there is no need to keep track of the actual value of the record at a given time step. The knowledge of the actual value of the record at a given time step is however required for bridges, where the random walk returns back to the origin after $n$ time steps. This is done here by considering the full joint distribution of the ages $\tau_i$'s and the record increments $\rho_i$'s (which are the differences between two consecutive records), see Fig. \ref{Fig1}. The computation of this joint distribution is the main technical achievement of the present paper [see Eq. (\ref{eq:def_joint_pdf_exp_2}) below]. Finally, yet another noticeable difference between free RWs and RWs bridges, from the point of view of records, is that there is no simple relation between $Q^{\alpha}_{B}(n)$ and $\langle \ell^\alpha_{\max,B}(n)\rangle$ for bridges, while they are directly related for free random walks (\ref{eq:relQ_lmax}). Hence, one expects that $Q^{\alpha}_{B}(n)$ and $\langle \ell^\alpha_{\max,B}(n)\rangle$ generically lead to two different universal constants, which we can compute explicitly in the case of the discrete $\alpha = {\rmd}$ and exponential $\alpha = \e$ distributions. 

Our main results can be summarized as follows (see also Table \ref{table:summary}). First, for discrete random walk (\ref{def_discrete}) we obtain exact results for the full distribution of the records $P_B^{\rmd}(m,n) = \Pr(R_B^{\rmd}(n) = m,n)$. In particular, for large $n$ (and $n$ even as a discrete random walk bridge has necessarily an even number of steps), we show that
\beq\label{eq:summary_av_R_discrete}
 \langle R_B^{\rmd}(n) \rangle \sim A^{\rmd}_B \sqrt{n} \;, \; A^{\rmd}_B = \frac{\sqrt{\pi}}{2^{3/2}} \;.
\eeq
Hence it also grows like $\sqrt{n}$, as for free RW (\ref{eq:av_records_free_dicrete}), but a with a different prefactor: $A_B^\rmd/A^\rmd = {\pi}/{4} < 1$. On the other hand, for large $n$ and large $m$, keeping $X = m/\sqrt{n}$ fixed, the distribution $P_B^{\rmd}(m,n)$ takes the scaling form 
\beq\label{eq:summary_pdf_discrete}
P_B^{\rmd}(m,n) \sim \frac{1}{\sqrt{n}} \varphi_B^{\rm d}\left(X = \frac{m}{\sqrt{n}} \right) \;, \; \varphi_B^{\rmd}(X) = 4 \, X \, \e^{-2 X^2} \;, \; X > 0 \;,
\eeq
which is different from its counterpart for free RW (\ref{eq:distrib_R_free_discrete}). 

For continuous jump distribution $p^{\rm c}(\eta)$, we obtain the large $n$ behavior of $\langle R_B^{\rm c}(n)\rangle$ as
\beq\label{eq:av_records_bridge}
\langle R_B^{\rm c}(n)\rangle \sim A^{\rm c}_B(\mu) \sqrt{n} \;\;, {\rm as} \;\; n \to \infty \;,
\eeq
where $A_B^{\rm c}(\mu)$ is given in terms of an integral which depends explicitly on $\mu$ [see Eqs.~(\ref{eq:scaling_rate}) and (\ref{eq:av_rn_levy})]. In the case $\mu = 2$, this amplitude can be evaluated explicitly as
\beq\label{eq:av_records_ampli_bridge}
A^{\rm c}_B(\mu = 2) = \frac{\sqrt{\pi}}{2} \;,
\eeq
which should be compared to the value of the amplitude $A^{\rm c} = 2/\sqrt{\pi}$, independently of $\mu$, for the free RWs (\ref{eq:av_records_free}), and hence $A_B^{\rm c}(\mu =2)/A^{\rm c} = {\pi}/{4} < 1$. By comparing this result (\ref{eq:av_records_ampli_bridge}) with the one obtained for discrete random walks, we obtain that the ratio $A^{\rmd}_B/A^{\rm c}_B(\mu=2) = 1/\sqrt{2}$, as for free RWs. Although the analysis of the statistics of $R_B^{\rm c}(n)$, beyond the first moment, is quite difficult, it is possible to compute exactly the full distribution $P_B^{\e}(m,n)$ of the number of records $R^{\rm e}_B(n)$, for any finite $n$, for the case of the exponential distribution $p^{\e}(\eta)$ (\ref{def_exp}), which is representative of the case $\mu = 2$ [see Eq. (\ref{def_p_c})]. In particular, 
for large $n$ and large $m$, keeping $X = m/\sqrt{n}$ fixed, the distribution $P_B^{\e}(m,n)$ takes the scaling form 
\beq\label{eq:summary_pdf_exp}
P_B^{\e}(m,n) \sim \frac{1}{\sqrt{n}} \varphi_B^{\rm e}\left(X = \frac{m}{\sqrt{n}} \right) \;, \; \varphi_B^{\e}(X) = 2\, X \, \e^{-X^2} \;, \; X > 0 \;,
\eeq
which is different from its counterpart for free RW (\ref{eq:distrib_R_free}), and also slightly different from the corresponding distribution $\varphi_B^{\rm d}(X)$ for the discrete bridges (\ref{eq:summary_pdf_discrete}). 

On the other hand, for the record breaking probability $Q_B^{\alpha}(n)$, we obtain exact results for the discrete ($\alpha = {\rmd}$) and exponential ($\alpha = \e$) jump distributions. In particular, for large $n$, we show that these quantities converge to the same constant which can be computed exactly in terms of a single integral given in Eq. (\ref{eq:cQ}). This integral can be easily evaluated numerically with arbitrary accuracy, yielding
\beq\label{eq:intro_qinf_bridge}
&&\lim_{n \to \infty} Q_B^{\rm d}(n) = \lim_{n \to \infty} Q_B^{\rm e}(n) = Q_B^{\rmd}(\infty) = Q_B^\e(\infty) = 0.6543037\ldots \;,
\eeq
which is different from the one characterizing free RWs (\ref{eq:qinf_free}), $Q_B^{\rmd}(\infty) > Q^{\rmd}(\infty)$. Furthermore, for the discrete and exponential distributions, we compute exactly the average age of the longest lasting record $\langle \ell_{\max,B}^{\alpha}(n)\rangle$ and show that
\beq
\hspace*{-1cm}\lim_{n \to \infty} \frac{\langle \ell_{\max,B}^{\rmd}(n)\rangle}{n} = \lim_{n \to \infty} \frac{\langle \ell_{\max,B}^{\e}(n)\rangle}{n} = \lambda_{\max,B}^{\rmd}= \lambda_{\max B}^{\e} = 0.6380640\ldots \label{eq:intro_const_lmax_exp}
\eeq 
where, again, this constant can be expressed in terms of an integral given in Eq. (\ref{eq:const_lmax_exp}). Note in particular that $\lambda^{\alpha}_{\max, B} \neq Q_B^{\alpha}(\infty)$, while the two corresponding quantities coincide for free RWs [see Eqs. (\ref{eq:asympt_lmax}), (\ref{eq:asympt_lmax_discrete})]. Although we can not prove it, we expect that our results obtained for the exponential case for the large $n$ behavior of the distribution of the number of records (\ref{eq:summary_pdf_exp}), the probability of the record breaking (\ref{eq:intro_qinf_bridge}) and the age of the longest lasting record (\ref{eq:intro_const_lmax_exp}) are valid for any continuous distribution $p^{\rm c}(\eta)$ with finite $\sigma$, i.e., with $\mu = 2$. This conjecture is corroborated by our numerical simulations of the record statistics of RW bridges with a Gaussian jump distribution.

The paper is organized as follows. In section \ref{sec:mean_R}, we compute exactly the mean number of records for RW bridges, $\langle R_B^{\alpha}(n)\rangle$, for the discrete and exponential distributions (for any value of $n$ in these cases) and for generic continuous distributions (in the large $n$ limit). In section \ref{sec:distrib_R} we compute exactly the full distribution of the number of records $R_B^\alpha(n)$ for the discrete and exponential distributions for arbitrary $n$. Section \ref{sec:ages} is devoted to the statistics of the ages of the records, where we compute $Q^{\alpha}_B(n)$ and $\langle \ell_{\max,B}^\alpha(n)\rangle$, again for the discrete and exponential jump distributions. In section \ref{sec:simu} we compare our exact analytical results to numerical simulations before we conclude in section \ref{sec:conclusion}. Some technical details have been left in the appendices A, B and C.

\section{Mean number of records}\label{sec:mean_R}

Let $R_B^\alpha(n)$ be the number of records up to step $n$ for a random walk bridge. We can then write
\beq\label{def_sigma}
R_B^\alpha(n) = \sum_{k=0}^{n} \sigma^\alpha(k,n) \;,
\eeq
where $\sigma^\alpha(k,n)$ is a binary random variable taking values 0 or 1. The variable $\sigma^\alpha(k,n) = 1$ if a record happens at step $k$ and $\sigma^\alpha(k,n)=0$ otherwise, for a random walker which comes back to the origin after $n$ steps. By convention $\sigma^\alpha(k=0,n) = 1$ as $x_B(0)=0$ is a record. On the other hand one has obviously $\sigma^\alpha(k,n) = 0$ for $k\geq n$ [in particular $\sigma^\alpha(n,n) = 0$ because 
$x_B(n) = x_B(0) = 0$ is never a record -- as a record is defined by a strict inequality in Eq. (\ref{def_record})]. Hence, the mean number of records up to step $n$ is given by:
\beq\label{av_record_sum}
\langle R_B^\alpha(n) \rangle = \sum_{k=0}^{n-1} \langle \sigma^\alpha(k,n) \rangle = \sum_{k=0}^{n-1} r^\alpha(k,n) \;,
\eeq
where $r^\alpha(k,n)=\langle \sigma^\alpha(k,n) \rangle$ is the record rate, i.e., the probability that a record happens at step $k$. Therefore, to compute $\langle R_B^\alpha(n) \rangle $, we will first evaluate $r^\alpha(k,n)$ and then sum over $k$ (\ref{av_record_sum}). For $n\geq 1$, the evaluation of $r^\alpha(k,n)$ relies on the two following quantities:
\begin{itemize}
\item[$\bullet$]{The free Green's function (propagator) $G^\alpha(x,x_0,n)$ that denotes the probability (discrete distribution) or probability density (continuous distribution) that a random walker starting at $x_0$ arrives at $x$ after $n$ steps.}
\item[$\bullet$]{The constrained Green's function $G^\alpha_>(x,x_0,n)$ that denotes the probability (discrete distribution) or probability density (continuous distribution) that a random walker starting at $x_0$ arrives at $x$ after $n$ steps and staying strictly positive in-between.} 
\end{itemize}
\begin{figure}
\centering
\includegraphics[width=0.7\linewidth]{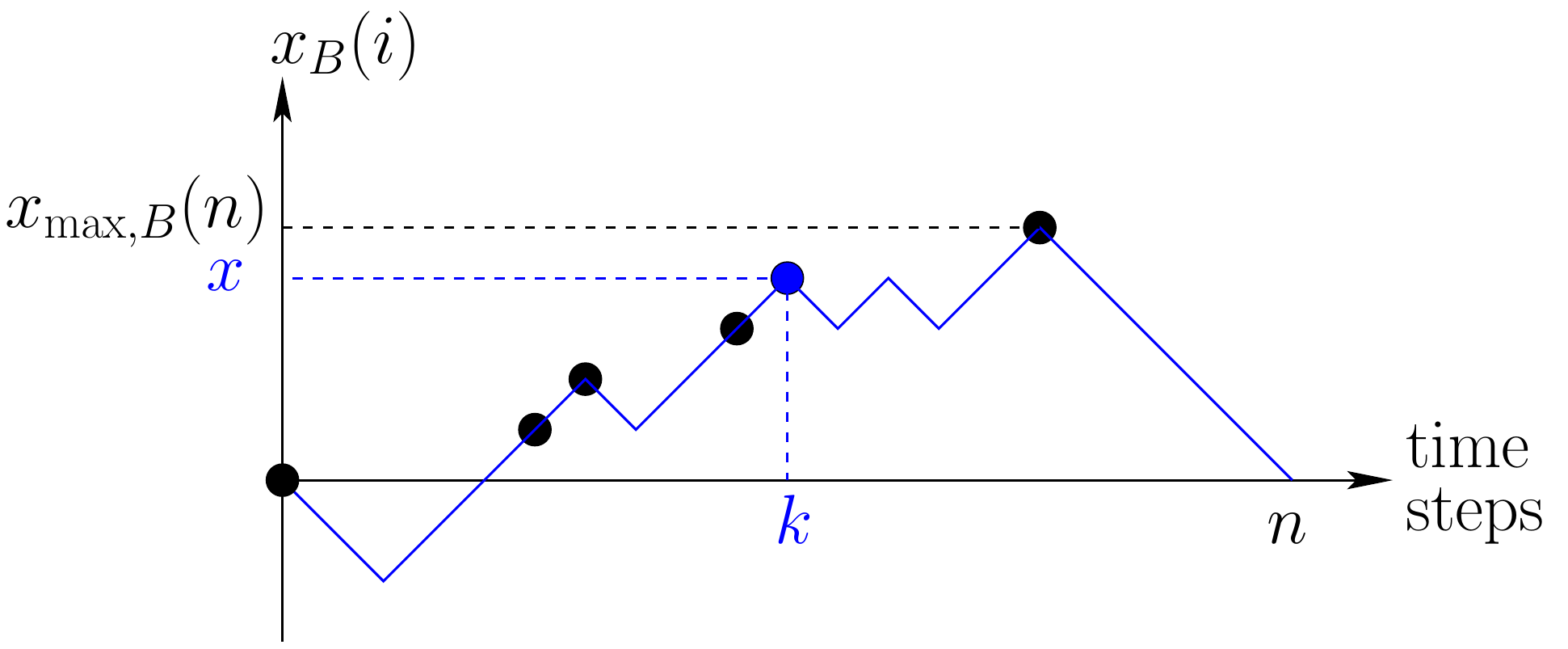}
\caption{A discrete random walk bridge of $n=20$ steps. Here the number of records is $R_{20}= x_{\max,B}(20)+1 = 6$.}\label{Fig2}
\end{figure}

To compute $r^\alpha(k,n)$, let us suppose that a record happens at step $k$ with a record value $x$ (see Fig. \ref{Fig2}). This corresponds to the event that the walker, starting at the origin at step $0$, has reached the level $x$ for the first time at step $k$ and returns back to the origin after $n$ steps -- as we are considering random walk bridges. In the time interval $[0,k]$, the walker propagates from $0$ to $x$, being constrained to stay {\it strictly} below $x$. To compute the corresponding propagator, we take $x$ as the new origin of space and then reverse both the time and coordinate axis. Hence, we see that on the time interval $[0,k]$, the particle propagates with $G^\alpha_{>}(x,0,k)$. On the other hand, between step $k$ and step $n$ (where the walker ends at the origin) the walker is free and thus propagates with $G^\alpha(0,x,n-k) = G^\alpha(x,0,n-k)$, as the jump distribution is symmetric. The record rate is then obtained by integrating the probability of this event over $x \geq 0$ as the record can take place at any level $x\geq 0$ (note that only the first record, i.e., $k=0$, is such that $x=0$). Using the statistical independence of the random walk in the time intervals $[0,k]$ and $[k,n]$ (being Markovian), one thus has, for $n \geq 1$:
\beq\label{expr_rm}
\hspace*{-1.5cm}r^\alpha(k,n) = \frac{1}{G^\alpha(0,0,n)}\int_0^\infty \, \rmd x \, G^\alpha_>(x,0,k) G^\alpha(x,0,n-k) \, \;, \; 0 \leq k \leq n-1 \;,
\eeq
where we have divided by $G^\alpha(0,0,n)$ as we are considering random walks that are conditioned to come back to the origin after $n$ time steps (bridges). Note that in the case of a discrete random walk, i.e., $\alpha = {\rmd}$, the integral over $x$ in Eq. (\ref{expr_rm}) has to be replaced by a discrete sum. Let us now analyze this formula (\ref{expr_rm}) for different types of random walks.

\subsection{Discrete random walk}

In the case of a discrete random walk, it is convenient to compute $\langle R_B^{\rmd}(n) \rangle = \sum_{k=0}^n r^{\rmd}(k,n)$ as
\begin{eqnarray}\label{eq:Rn_discrete}
\langle R^{\rm d}_B(n) \rangle = \frac{\langle R^{\rm d}_B(n) \rangle_{(0)} }{G^{\rm d}(0,0,n)}
\end{eqnarray}
where, here and in the following, the subscript `$(0)$' refers to a random walk starting from $x(0)=0$ and ending at $x(n)=0$ after $n$ steps and $G^{\rmd}(0,0,n)$ is the probability for this event. The generating function of the numerator $\langle R^{\rm d}_B(n) \rangle_{(0)}$ is from Eq. (\ref{expr_rm}) 
\beq\label{gf_rn_discrete}
\sum_{n=0}^\infty z^n \langle R^{\rm d}_B(n) \rangle_{(0)}= \sum_{x=0}^\infty \tilde G^{\rm d}_>(x,0,z) \tilde G^{\rm d}(x,0,z) \;,
\eeq
where the propagators can be explicitly computed (see \ref{sec:appendix_discrete}):
\beq\label{G_discrete}
\tilde G^{\rm d}(x,0,z) = \sum_{n=0}^\infty z^n G^{\rm d}(x,0,n) = \frac{1}{\sqrt{1-z^2}} \left(\frac{1-\sqrt{1-z^2}}{z} \right)^x
\eeq
and 
\beq\label{G>discrete}
\tilde G^{\rm d}_>(x,0,z) = \sum_{n=0}^\infty z^n G_>^{\rm d}(x,0,n) = \left(\frac{1-\sqrt{1-z^2}}{z} \right)^{x} \;.
\eeq
Hence one obtains by summing over $x$ in Eq. (\ref{gf_rn_discrete}), using Eqs. (\ref{G_discrete}) and (\ref{G>discrete})
\beq\label{eq:gf_Nn_discrete}
\sum_{n=0}^\infty z^n \langle R^{\rm d}_B(n) \rangle_{(0)} = \frac{1}{2(1-z^2)} + \frac{1}{2\sqrt{1-z^2}} \;.
\eeq
Therefore one obtains
\beq\label{eq:Nn_discrete}
\langle R^{\rm d}_B(n) \rangle_{(0)} = 
\begin{cases}
& \dfrac{1}{2} + \dfrac{1}{2^{2k+1}}{2k \choose k} \;, \; {\rm for}\; n \; {\rm even}\;, n = 2k\\
& 0 \;, \; {\rm for \;} n \; {\rm odd} \;
\end{cases} \;.
\eeq
Using that the denominator in Eq. (\ref{eq:Rn_discrete}) is given by $G^{\rmd}(0,0,n) = 2^{-n} {n \choose n/2}$ if $n$ is even and 0 otherwise one gets finally 
\begin{eqnarray}\label{eq:exact_av_Rn_discrete}
\langle R_B^{\rmd}(n) \rangle = 
\begin{cases}
&2^{n-1} {n \choose n/2}^{-1} + \frac{1}{2} \;, \, n \; {\rm even} \;, \\
&0 \;, \, n {\; \rm odd} \;.
\end{cases}
\end{eqnarray}
The first terms are $\langle R_B^{\rm d}(0) \rangle = 1$, $\langle R_B^{\rm d}(2) \rangle = 3/2$, $\langle R_B^{\rm d}(4) \rangle = 11/6$. Finally, for large $n$ it behaves like
\begin{eqnarray}\label{eq:asympt_av_Rn_discrete}
\langle R_B^{\rm d}(n) \rangle \sim 
\begin{cases}
&\dfrac{\sqrt{\pi}}{2^{3/2}} \sqrt{n} \;, \, {\rm for} \; n \; {\rm even} \;, \\
&0 \;, \, {\rm for} \; n {\; \rm odd} \;,
\end{cases}
\end{eqnarray}
as announced in Eq. (\ref{eq:summary_av_R_discrete}).

\subsection{Exponential jump distribution}

In this case, the jumps $\eta_i$'s are continuous variables with a distribution given by $p^\e(\eta) = \frac{1}{2b} \e^{-|\eta|/b}$. As we will see, the record statistics for the exponential distribution is exactly solvable. Indeed, in this case, both the free and the constrained Green's functions can be computed exactly \cite{CM05}. The associated generating functions (GFs) $\tilde G^\e(x,0,z)$ (free) and  $\tilde G^\e_>(x,0,z)$ (constrained) read \cite{CM05} (see also \ref{sec:appendix_exp})
\begin{eqnarray}
&&\tilde G^\e(x,0,z) = \sum_{n=1}^\infty z^n G^\e(x,0,n) =  \frac{z}{2b\sqrt{1-z}} \e^{-\frac{|x|}{b}\sqrt{1-z}} \label{eq:expr_GG>_exp} \\
&&\tilde G^\e_>(x,0,z) = \sum_{n=0}^\infty z^n G^\e_>(x,0,n) = \delta(x) + \frac{1-\sqrt{1-z}}{b} \e^{-\frac{|x|}{b}\sqrt{1-z}} \;. \label{eq:expr_GG>_exp2}
\end{eqnarray}
Note that the definition of the GF of the free propagator $\tilde G^\e(x,0,z)$ in Eq. (\ref{eq:expr_GG>_exp}) does not contain the $n=0$ term
-- as it will not enter into the calculations -- while this term $n=0$ has to be retained for the constrained propagator and produces a delta function. 

The average number of records is given by $\langle R^\e_B(n)\rangle = \sum_{k=0}^{n-1} r^{\e}(k,n)$ where the records rates $r^{\e}(k,n)$ are given by Eq.~(\ref{expr_rm}). To proceed, we write as above in Eq.~(\ref{eq:Rn_discrete}):
\begin{eqnarray}\label{eq:Rn_exp}
\langle R_B^{\e}(n)\rangle = \frac{ \langle R_B^{\e}(n)\rangle_{(0)}}{G^{\e}(0,0,n)} \;, \; n \geq 1 \;,
\end{eqnarray}
where the numerator $\langle R_B^{\e}(n)\rangle_{(0)}$ is given by
\begin{eqnarray}
\langle R_B^{\e}(n)\rangle_{(0)} = \sum_{k=0}^{n-1} \int_0^\infty \, \rmd x \, G^{\rm e}_>(x,0,k) G^{\rm e}(x,0,n-k) \;, \; n \geq 1.
\end{eqnarray}
Hence its GF is given by
\begin{eqnarray}
&&\hspace*{-1.5cm}\sum_{n=1}^\infty z^n \langle R_B^{\e}(n)\rangle_{(0)} = \int_0^\infty \rmd x \, \tilde G^{\e}_>(x,0,z) \tilde G^\e(x,0,z) = \frac{1}{4b} \left(\frac{z}{1-z} + \frac{z}{\sqrt{1-z}} \right)\label{eq:expr_gf_deno_exp1} \;,
\end{eqnarray}
where in the last equality we have used the explicit expressions of $\tilde G^\e(x,0,z)$ and $\tilde G^{\e}_>(x,0,z)$ given in Eqs. (\ref{eq:expr_GG>_exp}) and (\ref{eq:expr_GG>_exp2}). From Eq. (\ref{eq:expr_gf_deno_exp1}) we easily extract $\langle R_B^{\e}(n)\rangle_{(0)}$ for all $n$. Finally, from Eq. (\ref{eq:Rn_exp}) one obtains the mean number of records as
\beq\label{eq:exact_av_Rn_exp}
\langle R_B^\e(0) \rangle = 1 \;, \langle R_B^\e(n) \rangle = \frac{1}{2} + \frac{2^{2n-3}}{{2n-2 \choose n-1}} \; {\rm for\;} n \geq 1 \;,
\eeq
where we have used $G^\e(0,0,n) = {2n-2 \choose n-1}/(b 2^{2n-1})$, for $n \geq 1$. Interestingly, by comparing Eqs. (\ref{eq:exact_av_Rn_discrete}) and (\ref{eq:exact_av_Rn_exp}) one sees that $\langle R^{\rm e}_B (n)\rangle = \langle R_B^\e(2n-2)\rangle$. For large $n$, one finds straightforwardly from Eq. (\ref{eq:exact_av_Rn_exp}) 
\beq\label{eq:asympt_av_Rn_exp}
\langle R_B^\e(n) \rangle \sim \frac{\sqrt{\pi}}{2} \sqrt{n} \;,
\eeq
which is $\sqrt{2}$ times the result obtained for the discrete random walk in Eq.~(\ref{eq:asympt_av_Rn_discrete}).

\subsection{Stable jump distribution}

Now we consider the more general situation where the jumps are continuous random variables distributed according to $p^{\rm c}(\eta)$ as given in Eq.~(\ref{def_p_c}). Although an exact calculation of $\langle R^{\rm c}_B(n) \rangle$ for any finite $n$ seems quite difficult in this case, one can perform a large $n$ asymptotic analysis as follows. 

We recall that the average number of records is given by $\langle R_B^{\rm c}(n) \rangle = \sum_{k=0}^n r^{\rm c}(k,n)$. One can show that this sum over $k$ is dominated by the values of $k \sim {\cal O}(n)$ which are thus large, when $n \gg 1$. Hence, to evaluate the record rates $r^{\rm c}(k,n)$ given in Eq.~(\ref{expr_rm}) for large $k$ one can replace the propagators $G^{\rm c}(x,0,n-k)$ and $G^{\rm c}_>(x,0,k)$ by their scaling forms valid for $k,n \gg 1$, with $k/n$ fixed, and $x \gg 1$, with $x/n^{1/\mu}$ fixed. One has indeed
\begin{eqnarray}
&&G^{\rm c}(x,0,n-k) \sim \frac{1}{a (n-k)^{1/\mu}} R\left(\frac{x}{a \, (n-k)^{1/\mu}} \right)\;, \label{eq:G_scaling} \\
&&G^{\rm c}_>(x,0,k) \sim \frac{1}{a \sqrt{\pi}k^{1/2+1/\mu}} R_+ \left(\frac{x}{a \, k^{1/\mu}} \right) \;, \label{eq:G>_scaling}
\end{eqnarray} 
where the scaling functions are normalized, i.e., $\int_{-\infty}^\infty \rmd x \, R(x) = 1$ and $\int_0^\infty \rmd x \, R_+(x) = 1$. The scaling function $R(x)$ is a L\'evy stable distribution:
\beq\label{eq:stable_dist}
R(x) = \frac{1}{2 \pi} \int_{-\infty}^\infty \rmd k \, \e^{-i k x} \e^{-|k|^\mu} \;,
\eeq
and in particular $R(0) = {\Gamma(1+1/\mu)}/{\pi}$. On the other hand, there is no explicit expression for $R_+(x)$ for generic $\mu < 2$, while
for $\mu = 2$ one has $R_+(x) = 2 \,x \, \e^{-x^2}$. 

With such a normalization 
one can check in particular that by integrating $G^{\rm c}_>(x,0,k)$ in Eq. (\ref{eq:G>_scaling}) over $x$ one recovers the survival probability $q(k)$, which is the probability that the walker, starting from $x(0)=0$ stays positive up to step $k$:
\beq
\int_0^\infty {\rmd x} \; G^{\rm c}_>(x,0,k) = q(k) \sim \frac{1}{\sqrt{\pi \,k }} \;, \; {\rm as} \; k \to \infty \;,
\eeq
in agreement with the Sparre Andersen theorem \cite{SA53}. By inserting these scaling forms (\ref{eq:G_scaling}, \ref{eq:G>_scaling}) into the expression for $r^{\rm c}(k,n)$ in (\ref{expr_rm}) one finds, that for large $k$ and $n$ keeping $k/n = y$, fixed (with $0 \leq y \leq 1$): 
\beq\label{eq:scaling_rate}
&&r^{\rm c}(k,n) = \frac{1}{\sqrt{n}} H\left(y = \frac{k}{n}\right) \;, \nonumber \\
&&H(y) = \frac{\sqrt{\pi}}{\Gamma(1+1/\mu)} \frac{1}{\sqrt{y}(1-y)^{1/\mu}} \int_0^\infty \rmd x \, R_+(x) R\left(\frac{x}{(y^{-1}-1)^{1/\mu}}\right) \;.
\eeq
Finally, from this scaling form for the record rate (\ref{eq:scaling_rate}), one obtains finally
\begin{eqnarray}\label{eq:av_rn_levy}
\langle R_B^{\rm c}(n) \rangle = \sum_{k=0}^{n} r^{\rm c}(k,n) \sim A^{\rm c}_B(\mu )\sqrt{n} \;, \; A^{\rm c}_B(\mu) = \int_0^1 \rmd y \, H(y) \;.
\end{eqnarray} 
Hence $\langle R^{\rm c}_B(n) \rangle$ also grows like $\sqrt{n}$ for bridges but with an amplitude which depends on $\mu$, as announced in Eq.~ (\ref{eq:av_records_bridge}). In particular, one can check that $A^{\rm c}_B(\mu=2) = \sqrt{\pi}/2$, which coincides, as expected, with the result obtained in the exponential case, see Eq.~(\ref{eq:asympt_av_Rn_exp}).

\section{Full distribution of the number of records}\label{sec:distrib_R}

In the previous section, we computed the average number of records $\langle R^\alpha_B(n) \rangle$. Here we show that, in some cases, namely for discrete ($\alpha = {\rm d}$) and exponential ($\alpha = {\rm e}$) jump distributions, the full distribution of $R^\alpha_B(n)$ can be computed exactly, for any finite $n$.

\subsection{Discrete random walks}

In the case of discrete random walks, the distribution of $R_B^{\rmd}(n)$ can be computed by using the close relation between $R^{\rmd}_B(n)$ and the maximum of the random walk up to step $n$, denoted by $x_{\max,B}(n) = \max_{0\leq m\leq n} x_B(m)$ \cite{WMS2012} (see also Fig. \ref{Fig2}):
\begin{eqnarray}\label{eq:relation_nber_max}
R_B^{\rmd}(n) = x_{\max,B}(n) + 1 \;.
\end{eqnarray} 
To derive this relation (\ref{eq:relation_nber_max}), let us consider the time evolution of the two processes $R_B^{\rmd}(n)$ and $x_{\max,B}(n)$. At the next time step $n+1$, if a new site on the positive axis is visited for the first time, the process $x_{\max,B}(n)$ increases by 1, otherwise its value remains the unchanged. On the other hand, when this event happens, then the record number $R_B^{\rmd}(n)$ is also increased by 1, and otherwise it remains unchanged. Therefore we see that the two processes are locked with each other at all steps: for any realization of the random walk, we thus have $x_{\max,B}(n+1) - x_{\max,B}(n) = R_{B}^{\rmd}(n+1) - R_B^{\rmd}(n)$. Given that, initially one has $x_{\max,B}(0) = x_B(0) = 0$ and $R_B^{\rmd}(0) = 1$ (since, by convention, the first position is a record) one immediately obtains the relation in Eq. (\ref{eq:relation_nber_max}). Hence, this relation allows us to compute the PDF of $R_B^{\rmd}(n)$ as
\begin{eqnarray}
\hspace*{-2.5cm}P_B^{\rmd}(m,n) &=& \Pr (R_B^{\rmd}(n) = m) = \Pr (x_{\max,B}(n) = m-1) \\
\hspace*{-2.5cm}&=& \frac{P_B^{\rmd}(m,n)_{(0)}}{G^{\rm d}(0,0,n)} \;, \; P_B^{\rmd}(m,n)_{(0)}= \sum_{k=0}^{n} G^{\rmd}_>(m-1,0,k) G^{\rmd}_{\geq} (m-1,0,n-k) \label{eq:gf_dist_rn_discrete}
\end{eqnarray}
where $G^{\rmd}_{\geq}(x,x_0,k)$ is the probability that the random walker, starting at $x_0$, arrives at $x$ after $k$ steps, while staying non-negative (i.e., it may touch 0 but not $-1$) in-between. The generating function of the numerator $P_B^{\rmd}(m,n)_{(0)}$ in Eq. (\ref{eq:gf_dist_rn_discrete}) then reads
\beq\label{eq:gf_dist_rn_discrete2}
\sum_{n=0}^\infty P_B^{\rmd}(m,n)_{(0)} z^n = \tilde G^{\rmd}_>(m-1,0,z) \tilde G^{\rmd}_{\geq} (m-1,0,z) \;,
\eeq
where $\tilde G^{\rmd}_>(m-1,0,z)$ is given in Eq. (\ref{G>discrete}) and $\tilde G^{\rmd}_{\geq} (x,0,z)$ is given by (see \ref{sec:appendix_discrete})
\beq\label{Ggeqdiscrete}
\tilde G^{\rmd}_{\geq} (x,0,z) = \sum_{n=0}^\infty G^{\rmd}_{\geq} (x,0,n) z^n = \frac{2}{z} \left( \frac{1-\sqrt{1-z^2}}{z}\right)^{x+1} \;.
\eeq
Therefore Eq. (\ref{eq:gf_dist_rn_discrete2}) reads explicitly
\beq\label{eq:gf_dist_rn_discrete3}
\sum_{n=0}^\infty P_B^{\rmd}(m,n)_{(0)} z^n = \frac{2}{z} \left( \frac{1-\sqrt{1-z^2}}{z}\right)^{2m-1} \;.
\eeq

Here, to perform this computation, we show an alternative method which will also be useful to compute the statistics of the ages. To do so, we first introduce the ages of the records $\tau_i$'s and write the joint probability distribution of the $\tau_i$'s (see Fig. \ref{Fig1}) and the number of records $R_B^{\rmd}(n)$, 
\beq\label{joint_discrete}
&&\hspace*{-2.cm}P^{\rmd}(\ell_1, \cdots, \ell_{m-1}, a, m,n) = \Pr(\tau_1 = \ell_1, \ldots, \tau_{m-1} = \ell_{m-1}, A_n = a, R^{\rmd}_B(n) =m,n ) \nonumber \\
&&\hspace*{-2cm} = \frac{P^{\rmd}(\vec{\ell},m,n)_{(0)}}{G^{\rm d}(0,0,n)} \;, \; 
\eeq
where the numerator $P^{\rmd}(\vec{\ell},m,n)_{(0)}$ is given by
\beq\label{num_joint_discrete}
\hspace*{-2.cm} P^{\rmd}(\vec{\ell},m,n)_{(0)} = f^{\rmd}(\ell_1) \cdots f^{\rmd}(\ell_{m-1})G^{\rmd}_\geq(m-1,0,a) \delta\left(\sum_{i=1}^{m-1}\ell_i + a,n\right) \;,
\eeq
where $f^{\rmd}(\ell)$ is the first passage probability that the discrete RW, starting from $x_0$, arrives at $x_0+1$ for the first time at step $\ell$, and $\delta(a,b)$ denotes a delta Kronecker function. Since 
the RW is invariant under translation, this probability is independent of $x_0$ and for a discrete RW, its generating function is given by
\beq\label{eq:gf_first_passage_rw}
\tilde f^{\rmd}(z) = \sum_{\ell=1}^\infty f^{\rmd}(\ell) z^\ell = \frac{1-\sqrt{1-z^2}}{z} \;,
\eeq
from which we deduce that
\beq\label{eq:first_passage_discrete}
f^{\rmd}(\ell) = 
\begin{cases}
&0 \;, \; \ell \; {\rm even} \;, \\
& (-1)^{(\ell-1)/2} \dfrac{\sqrt{\pi}}{2 \Gamma(1-\ell/2)\Gamma(3/2+\ell/2)}\;, \ell \; {\rm odd} \;.
\end{cases}
\eeq

This joint distribution (\ref{joint_discrete}) contains all the information about the observables that we wish to compute here. In particular, the probability distribution of the record number $R_B^{\rmd}(n)$ is obtained by integrating it over the $\ell_i$'s:
\beq
\hspace*{-2cm}P_B^{\rmd}(m,n) = \sum_{\ell_1=1}^\infty \sum_{\ell_2=1}^\infty \ldots \sum_{\ell_{m-1}=1}^\infty \sum_{a=1}^\infty P^{\rmd}(\ell_1, \ell_2, \cdots, \ell_{m-1}, a,m,n) = \frac{P^{\rmd}_{B}(m,n)_{(0)}}{G^\rmd(0,0,n)}\,, \label{eq:gf_dist_rn_discrete_method2}
\eeq
where the generating function of the numerator is given by
\beq\label{eq:gf_dist_rn_discrete4}
\sum_{n=0}^\infty z^n P^{\rmd}_{B}(m,n)_{(0)} &=& [\tilde f^{\rmd}(z)]^{m-1} \tilde G^{\rmd}_{\geq}(m-1,0,z) \nonumber \\
&=&\frac{2}{z} \left( \frac{1-\sqrt{1-z^2}}{z}\right)^{2m-1} \;,
\eeq 
where we have used Eqs. (\ref{Ggeqdiscrete}) and (\ref{eq:gf_first_passage_rw}). Thus we see that this result (\ref{eq:gf_dist_rn_discrete4}) coincides with the one obtained previously by a quite different method (\ref{eq:gf_dist_rn_discrete3}). From this expression (\ref{eq:gf_dist_rn_discrete4}) we can compute $P^{\rmd}_{B}(m,n)_{(0)}$ (using Cauchy's formula) and finally $P^{\rmd}_{B}(m,n)$ as (for $n$ even)
\beq\label{eq:explicit_dist_rn_discrete}
\hspace*{-1.3cm}P^{\rmd}_{B}(m,n) = \frac{2^{n+1}}{{n \choose n/2}}\frac{(-1)^{n/2+m}}{(n/2+m)!} \sum_{j=0}^{2m-1} (-1)^j {2m-1 \choose j} \frac{\Gamma(j/2+1)}{\Gamma(j/2+1-n/2-m)} \;,
\eeq
for $1 \leq m \leq n/2+1$ (while $P^{\rmd}_{B}(m,n)=0$ for $m>n/2+1$, as the maximum value of a discrete RW bridge can not exceed $n/2$). Although the moments $\langle R_B^{\rmd}(n) \rangle$ can be obtained, in principle, from the full distribution in Eq. (\ref{eq:explicit_dist_rn_discrete}) it is more convenient to compute them from the GF in (\ref{eq:gf_dist_rn_discrete4}). In particular, one can easily check that one recovers the result for the average number of records $\langle R_B^{\rmd}(n) \rangle$ as obtained previously (\ref{eq:Rn_discrete}), (\ref{eq:gf_Nn_discrete}). Besides one can compute higher moments of $R_B^{\rmd}(n)$. For instance, the second moment is given by 
\begin{eqnarray}\label{eq:exact_var_Rn_discrete}
\langle \left[R_B^{\rmd}(n)\right]^2 \rangle = 
\begin{cases}
&\dfrac{n+1}{2} + \dfrac{\sqrt{\pi}}{2} \dfrac{\Gamma(n/2+1)}{\Gamma(n/2+1/2)}\;, \, n \; {\rm odd} \;, \\
&0 \;, \, n {\; \rm even} \;.
\end{cases}
\end{eqnarray}
For large $n$, it behaves as
\begin{eqnarray}\label{eq:asympt_var_Rn_discrete}
\langle \left[R_B^{\rmd}(n)\right]^2 \rangle \sim
\begin{cases}
&\dfrac{n}{2} \;, n \; {\rm odd} \;, \\
&0 \;, \, n {\; \rm even} \;.
\end{cases}
\end{eqnarray}
Finally, from Eq. (\ref{eq:gf_dist_rn_discrete3}), one can also extract the expression of $P_B^\rmd(m,n)$ in the large $n$ limit, which can be obtained by studying the limit $z \to 1$ of the GF in Eq. (\ref{eq:gf_dist_rn_discrete3}). In this limit, we set $z = \e^{-s}$ and investigate the limit $s \to 0$ where the discrete sum defining the GF in Eq. (\ref{eq:gf_dist_rn_discrete3}) can be replaced by an integral. Recalling that $P_B^\rmd(m,n) = 0$ if $n$ is odd, the left hand side of Eq. (\ref{eq:gf_dist_rn_discrete3}) can be written, in the limit $s \to 0$
\begin{eqnarray}
\sum_{n=0}^\infty P_B^{\rmd}(m,n)_{(0)} \e^{-s \,n} &=&  \sum_{k=0}^\infty P_B^{\rmd}(2\,k,n)_{(0)} \e^{-2s k} \nonumber \\
&\sim& \frac{1}{2} \int_0^\infty \rmd y \, \e^{- y\,s} P_B^{\rmd}(m,y)_{(0)} \;.
\end{eqnarray}
On the other hand, the right hand side of Eq. (\ref{eq:gf_dist_rn_discrete3}) assumes a simpler form in the limit $z = \e^{-s} \to 1$ such that Eq. (\ref{eq:gf_dist_rn_discrete3}) can be written
\beq\label{eq:laplace1}
\frac{1}{2} \int_0^\infty \rmd y \, \e^{- y\,s} P_B^{\rmd}(m,y)_{(0)} = 2 \, \e^{-2 \, m \sqrt{2 \, s}}. 
\eeq 
Hence $P_B^{\rmd}(m,y)_{(0)}$ can be straightforwardly obtained by Laplace inversion as
\beq
P_B^{\rmd}(m,n)_{(0)} \sim 4 \sqrt{\frac{2}{\pi}} \frac{m}{n^{3/2}} \e^{-2 m^2/n} \;, \; {\rm for} \; n \; {\rm even} \;.
\eeq
Finally, using that $G^{\rmd}(0,0,n) \sim \sqrt{2/\pi}\, n^{-1/2}$, one finds that the distribution $P_B^\rmd(m,n)$ takes the scaling form (valid for $n$ even):
\beq\label{eq:pdf_discrete}
P_B^{\rmd}(m,n) \sim \frac{1}{\sqrt{n}} \varphi_B^{\rm d}\left(X = \frac{m}{\sqrt{n}} \right) \;, \; \varphi_B^{\rmd}(X) = 4 X \e^{-2 X^2} \;,
\eeq
as announced in the introduction in Eq. (\ref{eq:summary_pdf_discrete}). Note that this distribution $\varphi_B^{\rmd}(X)$
coincides, up to a scale factor, with the probability distribution function of the maximum of a Brownian bridge on the unit time interval, as expected from the relation stated in Eq. (\ref{eq:relation_nber_max}). In particular, at variance with the result for the free random walk (\ref{eq:distrib_R_free_discrete}), this PDF is not a half-Gaussian distribution. In addition, it is easy to check that, from the moment of this distribution $\varphi_B^{\rmd}(X)$, one recovers the asymptotic result for $\langle R_B^{\rmd}(n) \rangle$ and $\langle \left[R_B^{\rmd}(n)\right]^2 \rangle$ obtained respectively in Eqs. (\ref{eq:asympt_av_Rn_discrete}) and~(\ref{eq:asympt_var_Rn_discrete}). 

\subsection{Exponential distribution}

For the exponential jump distribution $p^\e(\eta)$ (\ref{def_exp}), the starting point of our analysis is the equivalent of the joint distribution given, for discrete random walks, in Eq. (\ref{joint_discrete}). However, because $p^\e(\eta)$ is a continuous distribution, this computation is more delicate than in the discrete case. Indeed, as we are considering random walk bridges, the weight of the last part of the paths, where the walker comes back to origin, i.e., the last segment of duration $A_n = a$ (see Fig. \ref{Fig1}), involves the propagator $G^\e_{\geq}(Y,0,a) = G^\e_{>}(Y,0,a)$ [see Eq. (\ref{eq:gf_dist_rn_discrete})] where $Y=x_{\max,B}(n)$ is the actual value of the last record, which coincides with the maximum. For discrete random walk the number of records $R_B^{\rmd}(n)$ and $x_{\max,B}(n)$ are directly related through $x_{\max,B}(n)= R_B^{\rmd}(n)-1$ but this relation does not hold for a continuous jump distribution. Consequently, we need to keep track both of the number of records and of the value of the last record. A convenient way to do this is to introduce the record increments $\rho_i$'s and the joint distribution of the $\tau_i$'s, $\rho_i$'s, $R_B^\e(n)$ and $x_{\max,B}(n)$ (see Fig.~\ref{Fig1}):
\beq\label{eq:def_joint_pdf_exp}
&&\hspace*{-2.3cm}\Pr[\{\tau_i = \ell_i, \rho_i \in [r_i, r_i + \rmd r_i]\}_{1\leq i \leq m-1},A_n = a, R_B^\e(n)=m, x_{\max,B}(n) \in [Y,Y+\rmd Y],n] \nonumber \\
&& = P^{\e}(\{\ell_i, r_i\}_{1 \leq i \leq m-1},a,m,Y) \rmd r_1 \cdots \rmd r_{m-1} \rmd Y \;,
\eeq
where the joint PDF $P^\e(\{\ell_i, r_i\}_{1 \leq i \leq m-1},a,m,Y)$ is given by
\beq\label{eq:def_joint_pdf_exp_2}
\hspace*{-2cm} P^\e(\{\ell_i, r_i\}_{1 \leq i \leq m-1},a,m,Y) &=& \frac{1}{G^\e(0,0,n)}\prod_{i=1}^{m-1} \int_0^\infty \rmd y_i \, G^\e_>(y_i,0,\ell_i-1) p^{\e}(y_i + r_i) \nonumber \\ &&\times G^\e_>(Y,0,a) \, \delta\left(\sum_{i=1}^{m-1} r_i - Y\right) \delta\left(\sum_{i=1}^{m-1} \ell_i + a, n\right),
\eeq
where we have used that, for the exponential jump distribution which is continuous, $G^\e_{\geq}(x,0,n) = G^\e_>(x,0,n)$ for $x > 0$. In particular, the joint distribution of the $\tau_i$'s, $A_n$ and $R_B^{\e}(n)$, i.e., the equivalent of Eq. (\ref{joint_discrete}) for the discrete case, is obtained by integrating the formula in (\ref{eq:def_joint_pdf_exp}) over $r_i$'s and $Y$: 
\beq\label{eq:def_joint_pdf_exp_partial_int}
&&\hspace*{-2.3cm}P^{\e}(\ell_1, \ell_2, \cdots, \ell_{m-1}, a, m,n) = \Pr(\tau_1 = \ell_1, \ldots, \tau_{m-1} = \ell_{m-1}, A_n = a, R_B^\e(n) =m,n ) \nonumber \\
&&\hspace*{-2.3cm} = \frac{P^\e(\vec{\ell},m,n)_{(0)}}{G^\e(0,0,n)} \;, \; 
\eeq 
where 
\beq\label{eq:num_exp1}
&&\hspace*{-.cm}P^\e(\vec{\ell},m,n)_{(0)} = \prod_{i=1}^{m-1} \int_0^{\infty} \rmd r_i \, T(\ell_i,r_i) \nonumber \\
&&\times \int_0^\infty \rmd Y G^\e_>(Y,0,a) \, \delta\left(\sum_{i=1}^{m-1} r_i - Y\right) \delta\left(\sum_{i=1}^{m-1} \ell_i + a, n\right) \;,
\eeq
where
\beq\label{eq:expr_Tell}
T(\ell,r) = \int_0^\infty \rmd y \, G^\e_>(y,0,\ell-1) p^\e(y+r) \;.
\eeq
Note that this formula (\ref{eq:num_exp1}) together with (\ref{eq:expr_Tell}) is actually valid for any continuous jump distribution $p^{\rm c}(\eta)$ (\ref{def_p_c})-- where the superscript `e' is replaced by `c'. However, its analysis is in general very hard to do, mainly because the constrained propagator $G^{\rm c}_>(x,0,n)$ does not have any explicit expression, which prevents one to perform the analysis of this multiple integral. Fortunately, such an explicit expression exists for the case of an exponential jump distribution $p^\e(\eta) = 1/(2b) \e^{-|\eta|/b}$, which we now focus on.

In this case, the generating function of the constrained propagator $G^\e_>(x,0,n)$ is given by Eq. (\ref{eq:expr_GG>_exp2}), from which one gets that the GF of the building block $T(\ell,r)$ in Eq. (\ref{eq:expr_Tell}) is given by [using Eq. (\ref{eq:expr_GG>_exp2})]
\beq\label{eq:integral1}
\hspace*{-1.cm}\sum_{\ell = 1}^\infty T(\ell,r)\, z^{\ell} = z \int_0^\infty \rmd y \, \tilde G_>^\e(y,0,z) p^{\e}(y+r) = \frac{1}{b}(1-\sqrt{1-z}) \e^{-{r}/{b}} \;.
\eeq
From Eq. (\ref{eq:integral1}) one obtains
\beq\label{eq:expr_TGF}
T(\ell,r) = \frac{1}{b}f^{\rm e}(\ell) \, \e^{-r/b} \;, \; \sum_{\ell=1}^\infty f^{\rm e}(\ell) z^\ell = 1-\sqrt{1-z} \;,
\eeq
which yields the expression of the coefficients $f^{\rm e}(\ell)$ as
\beq\label{eq:expr_ck}
f^{\rm e}(\ell) = (-1)^{\ell +1} \frac{\sqrt{\pi}}{2\,\Gamma(3/2-\ell)\Gamma(\ell+1)} \sim \frac{1}{2\sqrt{\pi} \ell^{3/2}} \;, \; {\rm as} \; \ell \to \infty \;.
\eeq
By comparing this expression for $f^{\rm e}(\ell)$ and the one for $f^{\rm d}(\ell)$ in Eq. (\ref{eq:expr_ck}) we easily see that $f^{\rm e}(\ell) = f^{\rm d}(2\ell-1)$. By injecting this explicit expression of $T(\ell,r)$ (\ref{eq:expr_TGF}, \ref{eq:expr_ck}) in Eq.~(\ref{eq:num_exp1}), the joint probability distribution $P^\e(\vec{\ell},m,n)_{(0)}$ can be written
\beq
&&P^\e(\vec{\ell},m,n)_{(0)} = \prod_{i=1}^{m-1} f^{\e}(\ell_i) \int_0^\infty \rmd Y G^\e_>(Y,0,a) \, \e^{-Y/b} \nonumber \\
&&\times\prod_{i=1}^{m-1} \int_0^\infty \frac{\rm d r_i}{b}\, \delta\left(\sum_{i=1}^{m-1} r_i - Y\right) \delta\left(\sum_{i=1}^{m-1} \ell_i + a, n\right) \;.
\eeq 

Finally, using the identity
\beq\label{eq:identity}
\prod_{i=1}^{m-1} \int_0^\infty \rmd r_i \, \delta\left(\sum_{i=1}^{m-1} r_i - Y\right) = \frac{Y^{m-2}}{(m-2)!} \;,
\eeq
which can be easily shown by taking the Laplace transform on both sides of (\ref{eq:identity}) with respect to $Y$, we obtain an expression for the joint probability of the $\tau_i$'s, $A_n$ and $R_B^{\bf e}(n)$ as
\beq
&&P^\e(\vec{\ell},m,n)_{(0)} = \prod_{i=1}^{m-1} f^{\e}(\ell_i) q^{\rm e}(m,a) \delta\left(\sum_{i=1}^{m-1} \ell_i + a, n\right) \;, \label{eq:num_expfinal} \\
&&q^{\rm e}(m,a) = \frac{1}{(m-2)! b^{m-1}} \int_0^\infty \rmd Y \e^{-Y/b} \, Y^{m-2} \, G_>^{\rm e}(Y,0,a) \label{eq:qma_exp} \;,
\eeq
which has thus a structure very similar to the one found in the discrete case (\ref{num_joint_discrete}), but with different building blocks. Furthermore, the GF of $q^{\rm e}(m,a)$ in Eq. (\ref{eq:qma_exp}) can be obtained explicitly as (see \ref{sec:appendix_exp})
\beq\label{eq:GFq}
\tilde q^{\,\rm e}(m,z) = \sum_{a=1}^\infty q^{\rm e}(m,a) \, z^a = \frac{1}{b}\frac{1-\sqrt{1-z}}{(1+\sqrt{1-z})^{m-1}} = \frac{(1-\sqrt{1-z})^m}{b\, z^{m-1}} \;.
\eeq

From this joint distribution (\ref{eq:num_expfinal}), together with Eqs. (\ref{eq:expr_TGF}) and (\ref{eq:GFq}), it is possible to compute the statistics of all the observables which we want to study in this paper. In particular, one obtains the distribution of the number of records $P^\e_B(m,n)$ by summing it over $\ell_1, \ell_2, \ldots, \ell_{m-1}$ and $a$ as done before in the discrete case (\ref{eq:gf_dist_rn_discrete_method2}). This yields 
\beq\label{eq:gf_dist_rn_cont}
P_B^\e(m,n) = \frac{P_B^\e(m,n)_{(0)}}{G^\e(0,0,n)} \nonumber \\
\sum_{n=0}^\infty z^n P^\e_B(m,n)_{(0)} = [\tilde f^{\e}(z)]^{m-1} \tilde q^{\e}(m,z) &=& \frac{1}{b} \frac{(1-\sqrt{1-z})^m}{(1+\sqrt{1-z})^{m-1}} \nonumber \\
&=& \frac{1}{b} \frac{(1-\sqrt{1-z})^{2m-1}}{z^{m-1}} \;,
\eeq
from which one gets
\beq\label{eq:dist_rn_explicit}
\hspace*{-1.5cm}P_B^\e(m,n) = \frac{2^{2n-1}}{{2n-2 \choose n-1}} \frac{(-1)^{n+m-1}}{(n+m-1)!} \sum_{j=0}^{2m-1} (-1)^j {2m-1 \choose j} \frac{\Gamma(j/2+1)}{\Gamma(2+j/2-n-m)} \;,
\eeq
for $1 \leq m \leq n$, which is independent of the scale parameter $b$ [the factor $1/b$ in Eq. (\ref{eq:gf_dist_rn_cont}) is indeed cancelled by the denominator in the first line of Eq. (\ref{eq:gf_dist_rn_cont}), $G^{\rm e}(0,0,n) = {2n-2 \choose n-1}/(b \, 2^{2n-1})$]. By comparing this result for the exponential distribution (\ref{eq:dist_rn_explicit}) with the corresponding one for discrete jumps obtained before in Eq. (\ref{eq:explicit_dist_rn_discrete}), we easily find that $P^{\rm e}(m,n) = P^{\rmd}(m,2n-2)$, for $n \geq 1$. The moments of $R_B^\e(n)$ can be obtained, in principle, from the explicit expression of the full distribution in (\ref{eq:dist_rn_explicit}). It is however simpler to compute them from the GF in Eq. (\ref{eq:gf_dist_rn_cont}). For instance, from this expression, one can easily recover the value for $\langle R_B^\e(n)\rangle$ obtained by the previous method 
given in Eq.~(\ref{eq:exact_av_Rn_exp}). Furthermore, from this relation (\ref{eq:gf_dist_rn_cont}), one can also compute higher moments of $R_B^\e(n)$. In particular, the second moment is given by
\beq
\sum_{n=0}^\infty z^n \langle \left[R_B^\e(n)\right]^2\rangle_{(0)} = \frac{z}{4b(1-z)^{3/2}} + \frac{z}{4b(1-z)} 
\eeq
from which one obtains, for $n \geq 1$
\beq\label{eq:exact_var_Rn_exp}
\langle \left[R_B^\e(n)\right]^2\rangle = n - \frac{1}{2} + \frac{\sqrt{\pi}}{2} \frac{\Gamma(n)}{\Gamma(n-1/2)} \sim n \;, \; {\rm as} \; n \to \infty\;.
\eeq
Finally, from Eq. (\ref{eq:gf_dist_rn_cont}), one can obtain the limiting scaling form of the distribution of the number of records $P_B^\e(m,n)$ for $m \gg 1$, $n \gg 1$, keeping $m/\sqrt{n}$ fixed, as
\beq\label{eq:pdf_exp}
P_B^\e(m,n) \sim \frac{1}{\sqrt{n}} \varphi_B^\e\left(X = \frac{m}{\sqrt{n}}\right) \;, \; \varphi_B^\e(X) = 2 X \, \e^{-X^2} \;, \; X > 0\;,
\eeq
where we have used that $G^\e(0,0,n) \sim 1/(2b\sqrt{\pi n})$, for $n \gg 1$. This formula (\ref{eq:pdf_exp}) yields the result announced in the introduction in Eq. (\ref{eq:summary_pdf_exp}). Here also, one can easily check that, from the moments of the distribution $\varphi_B^\e(X)$ in Eq. (\ref{eq:pdf_exp}), one recovers the large $n$ behavior of $\langle R_B^\e(n)\rangle$ and $\langle \left[R_B^\e(n)\right]^2\rangle$ obtained respectively in Eqs. (\ref{eq:asympt_av_Rn_exp}) and (\ref{eq:exact_var_Rn_exp}).

\section{Statistics of the ages of records}\label{sec:ages}

In this section, we study in detail two quantities associated to the ages of the records: (i) the probability $Q_B^\alpha(n)$ (\ref{eq:def_Q}) that the age of the last record is the longest one for a random walk bridge and (ii) the statistics of $\ell^\alpha_{\max,B}(n)$ (\ref{eq:def_lmax}), which is the age of the longest lasting record. As in the previous section, we focus on the discrete ($\alpha = \rmd$) and the exponential ($\alpha = \e$) distributions. 

\subsection{Probability of record breaking $Q_B^\alpha(n)$}

The computation of $Q_B^\alpha(n)$ for a random walk bridge and for arbitrary jump distribution $p(\eta)$ is in general a very hard task. We show here that it can be computed exactly for the two special cases studied above: the discrete random walk ($\alpha = {\rmd}$) and the case of exponential jump distribution ($\alpha = \e$). 

\subsubsection{Discrete random walk.}

The probability $Q_B^{\rmd}(n)$ (\ref{eq:def_Q}) can be simply computed from the full joint PDF of the intervals given in Eq. (\ref{joint_discrete}) by summing it over the appropriate variables. For given values of the number of records $m$ and of $A_n = a$, we have to sum the joint PDF over 
the variables $\ell_i$'s from $1$ to $a$ -- as $A_n = a$ is the longest interval (see Fig. \ref{Fig1}). Finally, we sum over all the possible values of $a \geq 1$ over the number of records, for $m\geq 1$. 
\beq\label{eq:start_Q_discrete}
Q_B^{\rmd}(n) = \sum_{m=1}^\infty \sum_{a =1}^\infty \sum_{\ell_1=1}^a \cdots \sum_{\ell_{m-1}=1}^{a} P^{\rmd}(\vec{\ell}, m,n) \;,
\eeq
where $P^{\rmd}(\vec{\ell}, m,n)$ is explicitly given in Eq. (\ref{joint_discrete}) and (\ref{num_joint_discrete}). Again, we separate the numerator and the denominator and write
\beq\label{eq:num_Q_discrete}
&&Q_B^{\rmd}(n) = \frac{Q^{\rmd}_B(n)_{(0)}}{G^{\rmd}(0,0,n)} \nonumber \\
&& \sum_{n=0}^\infty Q^{\rmd}_B(n)_{(0)} z^n = \sum_{a=1}^\infty z^a \sum_{x=0}^a \left[h^{\rmd}(z,a)\right]^x G^{\rmd}_\geq(x,0,a) \;,
\eeq
where we have made the change of variable $x = m-1$ and where
\beq\label{eq:def_h}
h^{\rmd}(z,a) = \sum_{k=1}^a f^{\rmd}(k)z^k \;.
\eeq

From Eq. (\ref{eq:num_Q_discrete}) one can obtain, in principle, the value of $Q_B^{\rmd}(n)_{(0)}$ using Eq. (\ref{eq:def_h}) together with the explicit expression of $G^{\rmd}_\geq(x,0,a) $ given in Eq. (\ref{eq:ggeq_app}) -- for instance using Mathematica -- though obtaining
a closed form expression for $Q_{B}^{\rmd}(n)_{(0)}$ for any $n$ seems quite difficult. It is however possible to extract the large $n$ asymptotic behavior $Q_B^{\rmd}(n)_{(0)}$ by analyzing Eq. (\ref{eq:num_Q_discrete}) in the limit $z = \e^{-s}$ when $z \to 1$, which thus corresponds to $s \to 0$. In this limit, the double sum on the right hand side of Eq. (\ref{eq:num_Q_discrete}) is dominated by large $a$ and large $x$. In this limit, keeping $x/\sqrt{a}$ fixed, $G^{\rmd}_\geq(x,0,a)$ admits the following scaling form (for $x+a$ even):
\beq\label{eq:scaling_Ggeq}
G^{\rmd}_\geq(x,0,a) \sim 2\sqrt{\frac{2}{\pi a}} \frac{1}{\sqrt{a}} g^{\rmd}\left(\frac{x}{\sqrt{a}}\right) \;, \; g^{\rmd}(y) = y \, \e^{-\frac{y^2}{2}} \;,
\eeq 
while $G^{\rmd}_\geq(x,0,a) = 0$ for $x+a$ odd. Similarly, using the expression of $f^{\rmd}(\ell)$ given in Eq.~(\ref{eq:first_passage_discrete}), we obtain that in the scaling limit $a\to \infty$, $s \to 0$ (i.e., $z = \e^{-s} \to 1$), keeping the product $s \, a$ fixed, $h^{\rm d}(z,a)$ in Eq. (\ref{eq:def_h}) takes the scaling form
\beq\label{eq:scaling_h_discrete}
h^{\rmd}(z,a) \sim 1 - \sqrt{2s} \, F(a s) \;, \; F(y) &=& 1+ \frac{1}{2\sqrt{\pi}}\int_y^\infty \frac{\rmd u}{u^{3/2}} \e^{-u} \\
&=& {\rm erf}(\sqrt{y}) + \frac{1}{\sqrt{\pi}} \frac{\e^{-y}}{\sqrt{y}} \;. \nonumber
\eeq
From these asymptotic behaviors in Eqs. (\ref{eq:scaling_Ggeq}) and (\ref{eq:scaling_h_discrete}) one obtains that the sum over $x$ in Eq. (\ref{eq:num_Q_discrete}), which becomes an integral in the limit $s \to 0$, $a \to \infty$ keeping $y = a s$ fixed, takes the scaling form (as a function of the scaling variable $y$)
\beq
\sum_{x=0}^a \left[h^{\rmd}(z,a)\right]^x G^{\rmd}_\geq(x,0,a) &\sim& 2\sqrt{\frac{2}{\pi a^2}}\sum_{x=0}^a [1 - \sqrt{2s} F(a\,s)]^x g^{\rmd}\left(\frac{x}{\sqrt{a}} \right) \label{eq:asympt_suma} \\
&\sim& \sqrt{\frac{2}{\pi a^2}} \int_0^a \rmd x \, \e^{-x \sqrt{2s} F(a \,s)} g^{\rmd}\left(\frac{x}{\sqrt{a}} \right) \nonumber \\
&\sim& \sqrt{\frac{2}{\pi a}} G(y = a \, s) \;. \nonumber 
\eeq
Note that in the discrete sum over $x$ in the first line of Eq. (\ref{eq:asympt_suma}) only the terms such that $x+a$ is even contribute -- while the terms such that $x+a$ is odd are just zero. Hence this discrete sum is approximated by $1/2$ times the integral over $x$, as indicated in the second line of Eq. (\ref{eq:asympt_suma}). The function $G(y)$ in the third line of Eq.~(\ref{eq:asympt_suma}) is given~by 
\begin{eqnarray}\label{eq:def_G}
G(y) = 1 - \sqrt{\pi y} \, F(y) \, \exp{\left[y F^2(y) \right]} {\rm erfc}\left[\sqrt{y} F(y) \right] \;.
\end{eqnarray} 
Finally, one obtains the asymptotic behavior of the generating function of $Q_B^{\rmd}(n)_{(0)}$ in Eq.~(\ref{eq:num_Q_discrete}), as $z\to 1$
\beq\label{eq:asympt_denQ_discrete}
\sum_{n=0}^\infty Q_B^{\rmd}(n)_{(0)} z^n \sim \frac{c^{\rmd}_Q}{\sqrt{1-z}} \;, c^{\rmd}_Q =\sqrt{\frac{2}{\pi}} \int_0^\infty \frac{\rmd y}{\sqrt{y}} \e^{-y} \, G(y) \;, 
\eeq
such that $Q^{\rmd}_{B}(n)_{(0)} \sim 2\,c^{\rmd}_Q/\sqrt{\pi\,n}$ as $n \to \infty$, for $n$ even (while $Q_B^{\rm d}(n)_{(0)} = 0$ if $n$ is odd). Using that $G^{\rmd}(0,0,n) = 2^{-n} {n \choose n/2} \sim \sqrt{2/\pi}\, n^{-{1/2}}$, for $n \gg 1$, one obtains from Eq. (\ref{eq:start_Q_discrete}) together with Eq. (\ref{eq:asympt_denQ_discrete})
\beq\label{eq:cQ}
\hspace*{-1cm}\lim_{n \to \infty} Q_B^{\rmd}(n) = Q^{\rmd}_B(\infty) = \sqrt{2} \, c^{\rm d}_Q = \frac{2}{\sqrt{\pi}} \int_0^\infty \frac{\rmd y}{\sqrt{y}} \e^{-y} \, G(y) = 0.6543037\ldots \;,
\eeq
which is different from the corresponding value for free random walks, $Q^{\rmd}(\infty) = 0.626508 \ldots$ [see Eq. (\ref{eq:qinf_free_discrete})].

\subsubsection{Exponential jump distribution}

We now turn to the computation of $Q_B^\e(n)$ in Eq.~(\ref{eq:def_Q}) in the case of an exponential jump distribution. In this case, our starting point is the joint distribution of the ages of the records given in 
Eq. (\ref{eq:def_joint_pdf_exp_partial_int}). As before in the case of discrete random walks (\ref{eq:start_Q_discrete}), one has
 \beq\label{eq:start_Q_exp}
Q_B^\e(n) = \sum_{m=1}^\infty \sum_{a=1}^\infty \sum_{\ell_1=1}^a \cdots \sum_{\ell_{m-1}=1}^{a} P^\e(\vec{\ell}, m,n) \;,
\eeq
from which, using Eqs. (\ref{eq:def_joint_pdf_exp_partial_int}) and (\ref{eq:num_expfinal}), one obtains [similarly to the discrete case in Eq. (\ref{eq:num_Q_discrete})]
\beq\label{eq:num_Q_exp}
&&\hspace{-0cm}Q_B^\e(n) = \frac{Q^\e_B(n)_{(0)}}{G^\e(0,0,n)} \nonumber \\
&&\hspace*{-0cm}\sum_{n=0}^\infty Q_B^\e(n)_{(0)} z^n = \sum_{a=1}^\infty z^a \sum_{m=1}^\infty [h^{\e}(z,a)]^{m-1} q^\e(m,a) \;,
\eeq
where $f^\e(\ell)$ is given in Eq. (\ref{eq:expr_ck}).

To obtain the large $n$ behavior of $Q_B^\e(n)_{(0)}$, we need to analyze its generating function in Eq. (\ref{eq:num_Q_exp}) in the limit $z = \e^{-s} \to 1$, i.e. $s\to 0$. In this limit, the sum over $a$ is dominated by large values of $a \sim 1/s$. In this limit, $h^\e(z,a)$ takes the scaling form, for $s \to 0$, $a \, s$ fixed
\beq\label{eq:scaling_h_exp}
h^\e(z,a) \sim \left(1 - \sqrt{s} F(a \, s) \right) \;,
\eeq
where the scaling function $F(y)$ is given in Eq. (\ref{eq:scaling_h_discrete}). Furthermore, in the limit $z = \e^{-s} \to 1$ (i.e., $s \to 0$), the sum over $m$ in (\ref{eq:num_Q_exp}) is dominated by large values of $m$. In the limit, $m \to \infty$, $a\to \infty$, keeping $m/\sqrt{a}$ fixed, we obtain, 
from the GF of $q^\e(m,a)$ given in Eq. (\ref{eq:GFq}), that it takes the form
\beq\label{eq:qe_asympt}
q^\e(m,a) \sim \frac{1}{2\,b\sqrt{\pi}} \frac{m}{a^{3/2}} \e^{-{m^2}/{(4a)}} \;.
\eeq 
Therefore, by injecting these asymptotic behaviors given in (\ref{eq:scaling_h_exp}) and in (\ref{eq:qe_asympt}) into Eq. (\ref{eq:num_Q_exp}), one obtains
\beq
&&\hspace*{-2cm}\sum_{n=0}^\infty Q_B^\e(n)_{(0)}\e^{-s \,n} \sim \frac{1}{2 \, b \sqrt{\pi}} \sum_{a=1}^\infty \frac{\e^{-s a}}{a^{3/2}} \sum_{m=1}^\infty m \left(1 - \sqrt{s} F(a \, s)\right)^{m-1} \e^{-m^2/(4a)} \;.
\eeq
The sum over $m$ can then be approximated by an integral, in the limit $s \to 0$ as follows
\beq\label{eq:sum_m}
&&\sum_{m=1}^\infty m \left(1 - \sqrt{s} F(a \, s)\right)^{m-1} \e^{-m^2/(4a)} \sim \sum_{m=1}^\infty m \, \e^{-(m-1)\sqrt{s} F(a s)} \e^{-m^2/(4a)} \nonumber \\
&&\sim \int_0^\infty \rmd m \, m \, \e^{-(m-1) \sqrt{s} F(a s) - m^2/(4 a)} = 2 \, a \, G(a\,s) \;,
\eeq
where the function $G(y)$ is defined in Eq. (\ref{eq:def_G}). Hence, finally, approximating the remaining sum over $a$ by an integral (in the limit $s \to 0$) and performing the change of variable $y=s \, a$ one obtains
\beq
\sum_{n=0}^\infty Q_B^\e(n)_{(0)} z^n \sim \frac{c^\e_Q}{\sqrt{1-z}} \;, c^\e_Q = \frac{1}{b \sqrt{\pi}} \int_0^\infty \rmd y \frac{\e^{-y}}{\sqrt{y}} G(y)\;,
\eeq
which implies that, for large $n$, $Q^\e_B(n)_{(0)} \sim {c^\e_Q}/{\sqrt{\pi n}}$. Hence, using that $G^\e(0,0,n) \sim 1/(2 b \sqrt{\pi n})$, for large $n$, one obtains from the first line of Eq. (\ref{eq:num_Q_exp}) that
\beq
\lim_{n \to \infty} Q_B^\e(n) = Q_B^\e(\infty) = Q_B^{\rmd}(\infty) \;,
\eeq
which is thus the same constant as the one appearing in the discrete case given in Eq.~(\ref{eq:cQ}).

\subsection{Age of the longest lasting record}

\subsubsection{Discrete random walk.}

The average value of $\langle \ell^{\rmd}_{\max,B}(n) \rangle$ defined in Eq.~(\ref{eq:def_lmax}) can be computed from its cumulative distribution as
follows
\beq
\langle \ell^{\rmd}_{\max,B}(n) \rangle = \sum_{\ell=0}^\infty (1 - F^{\rmd}(\ell,n)) \;, \; F^{\rmd}(\ell,n) = \Pr[\ell_{\max,B}^{\rmd}(n) \leq \ell] \;,
\eeq
where $F^{\rmd}(\ell,n)=\Pr[\ell^{\rmd}_{\max,B}(n) \leq \ell]$ is simply obtained by summing up the joint PDF in Eq. (\ref{joint_discrete}) over $\ell_1, \cdots, \ell_{m-1}$ and $a$ from $1$ to $\ell$ -- as all of them have to be smaller than $\ell$ -- and finally over the number of records $m$ as follows:
\beq
\hspace*{-2.5cm}F^{\rmd}(\ell,n)=\Pr[\ell^{\rmd}_{\max,B}(n) \leq \ell] = \sum_{m=1}^\infty \sum_{a=1}^\ell \sum_{\ell_1=1}^\ell \cdots \sum_{\ell_{m-1}=1}^\ell P^{\rmd}(\ell_1, \ell_2, \cdots, \ell_{m-1}, a, m,n) \;.
\eeq
Therefore, one has
\beq\label{eq:starting_lmax_discrete}
\langle \ell^{\rmd}_{\max,B}(n) \rangle = \frac{\langle \ell^{\rmd}_{\max,B}(n) \rangle_{(0)}}{G^{\rmd}(0,0,n)} \;,
\eeq
where the numerator is given by
\begin{equation}
\hspace*{-0cm}\langle \ell^{\rmd}_{\max,B}(n) \rangle_{(0)} =  G^{\rm d}(0,0,n) - \sum_{m=1}^\infty \sum_{a=1}^\ell \sum_{\ell_1=1}^\ell \cdots \sum_{\ell_{m-1}=1}^\ell P^{\rmd}(\ell_1, \ell_2, \cdots, \ell_{m-1}, a, m,n)_{(0)}
\end{equation}
Therefore its generating function is given by
\be\label{eq:expr_lmax_2}
\hspace*{-0.cm}\sum_{n=0}^\infty \langle \ell^{\rmd}_{\max,B}(n) \rangle_{(0)}z^n = \sum_{\ell=0}^\infty \left(\tilde G^{\rmd}(0,0,z) - \sum_{a=1}^\ell z^a \sum_{m=1}^{a+1} \left[h^{\rmd}(z,\ell) \right]^{m-1} G^{\rmd}_{\geq}(m-1,0,a) \right) \,,
\ee
where the function $h^{\rmd}(z,\ell)$ is defined in Eq. (\ref{eq:def_h}). In the limit $z = \e^{-s} \to 1$, i.e. $s \to 0$, one can replace $G^{\rmd}_{\geq}(m-1,0,a)$ by its scaling form given in Eq. (\ref{eq:scaling_Ggeq}) and $h^{\rmd}(z,\ell)$ by its asymptotic behavior given in Eq. (\ref{eq:scaling_h_discrete}), in the scaling regime where $a$ and $\ell$ are large, keeping the products $s \, a$ and $s \, \ell$ fixed. The sum over $m$ in Eq. (\ref{eq:expr_lmax_2}) can then be analyzed along the same lines as done before in Eq. (\ref{eq:asympt_suma}) to yield:
\beq\label{eq:sum_m_lmax_discrete}
&&\hspace*{-.5cm}\sum_{m=1}^{a+1} \left[h^{\rmd}(z,\ell) \right]^{m-1} G^{\rmd}_{\geq}(m-1,0,a) \nonumber \\
&&\hspace*{-.5cm}\sim\sqrt{\frac{2}{\pi a}} \left[1 - \sqrt{\pi s a} \, F(s\, \ell) \, \exp{\left[s a F^2(s \ell) \right]} {\rm erfc}\left[\sqrt{s a} F(s \ell) \right] \right] \;, 
\eeq
where the function $F(y)$ is defined in Eq. (\ref{eq:scaling_h_discrete}). Inserting this asymptotic behavior (\ref{eq:sum_m_lmax_discrete}) in Eq. (\ref{eq:expr_lmax_2}), one finds
\beq\label{eq:gf_nlmax_discrete}
\sum_{n=0}^\infty \langle \ell^{\rmd}_{\max,B}(n) \rangle_{(0)} z^n\sim \frac{c_{\max}^{\rmd}}{(1-z)^{3/2}} \;, \; {\rm as} \; z \to 1\;,
\eeq
where the amplitude $c_{\max}^{\rmd}$ is given by [remembering that only the even values of $n$ contribute to the left hand side Eq. (\ref{eq:gf_nlmax_discrete})]
\beq\label{eq:def_C}
&&\hspace*{-0cm}c_{\max}^{\rmd} = \sqrt{2} \int_0^\infty \rmd t \left[\frac{1}{2} - I(t)\right],
\eeq
where the function $I(t)$ is given by
\beq\label{eq:def_I}
I(t) &=& \frac{1}{\sqrt{\pi}} \int_0^t \frac{\rmd y}{\sqrt{y}}\e^{-y}\left(1 - \sqrt{\pi y} F(t) \e^{y \, F^2(t)} {\rm erfc}\left(\sqrt{y} F(t) \right) \right) \\
&=& \frac{F(t)\e^{-t+t\,F^2(t)} {\rm erfc}[\sqrt{t} F(t)] -\e^{-t}/\sqrt{\pi t}}{1 - F^2(t)} \;.
\eeq
The behavior in Eq. (\ref{eq:def_C}) implies that $\langle \ell^{\rmd}_{\max,B}(n) \rangle \sim (4 c_{\max}^{\rmd}/\sqrt{\pi}) \sqrt{n}$, for large $n$ (even). Using finally that $G^{\rmd}(0,0,n) \sim \sqrt{2/(\pi\,n)}$ (again, for $n$ even) one finally obtains
\begin{eqnarray}
&&\hspace*{-0cm}\lim_{n \to \infty} \frac{\langle \ell^{\rmd}_{\max,B}(n)\rangle}{n} = \lambda^{\rmd}_{\max,B} = 2 \sqrt{2}\,c_{\max}^{\rmd} \\
&&\hspace*{-0cm} = 4 \int_0^\infty \rmd t \left(\frac{1}{2} -I(t)\right) \\
&&\hspace*{-0cm} = 0.6380640\ldots \label{eq:const_lmax_discrete} \;,
\end{eqnarray}
which is strictly smaller than the constant $Q^{\rmd}_B(\infty)$ given in Eq. (\ref{eq:cQ}).

\subsubsection{Exponential jump distribution.}

The average value of $\langle \ell^\e_{\max,B}(n) \rangle$ defined in Eq.~(\ref{eq:def_lmax}) can be computed from its cumulative distribution as
follows
\beq
\langle \ell^\e_{\max,B}(n) \rangle = \sum_{\ell=0}^\infty (1 - F^{\e}(\ell,n)) \;, F^{\e}(\ell,n)=\Pr[\ell^\e_{\max,B}(n) \leq \ell] \;,
\eeq
where $F^{\e}(\ell,n) = \Pr[\ell^\e_{\max,B}(n) \leq \ell]$ is simply obtained by summing up the joint PDF in Eq. (\ref{eq:def_joint_pdf_exp_partial_int}) over $\ell_1, \cdots, \ell_{m-1}$ and {$a$ from $1$ to $\ell$} and finally over the number of records $m$ as follows:
\beq
\hspace*{-2.5cm}F^{\e}(\ell,n)=\Pr[\ell^\e_{\max,B}(n) \leq \ell] = \sum_{m=1}^\infty \sum_{a=1}^\ell \sum_{\ell_1=1}^\ell \cdots \sum_{\ell_{m-1}=1}^\ell P^\e(\ell_1, \ell_2, \cdots, \ell_{m-1}, a, m,n) \;.
\eeq
Hence $\langle \ell^\e_{\max,B}(n) \rangle$ can be written as
\beq\label{eq:def_num_lmax_exp}
\hspace*{-2.5cm}&&\langle \ell^\e_{\max,B}(n) \rangle = \frac{\langle \ell^\e_{\max,B}(n) \rangle_{(0)}}{G^\e(0,0,n)} \;, \\
\hspace*{-2.5cm}&&\langle \ell^\e_{\max,B}(n) \rangle_{(0)} = \sum_{\ell=0}^\infty \left[G^\e(0,0,n) - \sum_{m=1}^\infty \sum_{a=1}^\ell \sum_{\ell_1=1}^\ell \cdots \sum_{\ell_{m-1}=1}^\ell P^\e(\ell_1, \cdots, \ell_{m-1},a,m,n)_{(0)}\right]. \nonumber
\eeq
Again, to extract the large $n$ behavior of $\langle \ell^\e_{\max,B}(n) \rangle_{(0)}$, it is convenient to analyze the generating function of $\langle \ell^\e_{\max,B}(n) \rangle_{(0)}$, which can be easily done by using the previous analysis. Indeed, one has
\beq\label{eq:gf_g00_exp}
\tilde G^\e(0,0,\e^{-s}) = \sum_{n=1}^\infty G^\e(0,0,n) \e^{- s n} \sim \frac{1}{2b} \frac{1}{\sqrt{s}} \;, \; {\rm as} \;\; s \to 0 \;.
\eeq
On the other hand, from the analysis performed above in Eqs. (\ref{eq:scaling_h_exp}, \ref{eq:qe_asympt}, \ref{eq:sum_m}) one has
\beq\label{eq:sum_m_lmnx_exp}
&&\sum_{n=1}^\infty \e^{-s n}\sum_{m=1}^\infty \sum_{\ell_1=1}^\ell \cdots \sum_{\ell_{m-1}=1}^\ell P^\e(\ell_1, \cdots, \ell_{m-1},a,m,n) \\
&&\sim \frac{1}{b \sqrt{\pi a}} \left[1 - \sqrt{\pi a s} \, F(s\, \ell) \, \exp{\left[a s F^2(s \ell) \right]} {\rm erfc}\left[\sqrt{s a} F(s \ell) \right] \right] \;,
\eeq
in the limit $s \to 0$, $a \to \infty$, $\ell \to \infty$, keeping $s \,a$ and $s \, \ell$ fixed and where the function $F(y)$ is defined in Eq.~(\ref{eq:scaling_h_discrete}). Therefore, using Eqs. (\ref{eq:gf_g00_exp}) and (\ref{eq:sum_m_lmnx_exp}), we obtain
\beq\label{eq:den_gf_exp}
\sum_{n=1}^\infty \langle \ell^\e_{\max,B}(n) \rangle_{(0)} z^n \sim \frac{c_{\max}^\e}{b (1-z)^{3/2}} \;, \; {\rm as} \; z \to 1 \;, 
\eeq
where the amplitude $c_{\max}^\e$ is given by
\beq
\hspace*{-0cm}c_{\max}^\e = \int_0^\infty \rmd t \left(\frac{1}{2} - I(t)\right) \;,
\eeq
where the function $I(t)$ is given in Eq. (\ref{eq:def_I}). From Eq. (\ref{eq:den_gf_exp}), one deduces that $\langle \ell^\e_{\max,B}(n) \rangle_{(0)} \sim (2 c_{\max}^\e/(b\,\sqrt{\pi})) \sqrt{n}$ for large $n$. Using that $G^\e(0,0,n) \sim 1/(2 b \sqrt{\pi n})$, for $n \gg 1$, we obtain finally
\begin{eqnarray}
&&\hspace*{-0cm}\lim_{n \to \infty} \frac{\langle \ell^\e_{\max,B}(n)\rangle}{n} = \lambda^\e_{\max,B} = 4\,c_{\max}^\e \\
&&\hspace*{-0cm} = 4 \int_0^\infty \rmd t \left(\frac{1}{2} - I(t)\right) \\
&&\hspace*{-0cm} = 0.6380640\ldots \label{eq:const_lmax_exp} \;,
\end{eqnarray}
which is exactly the same constant that appears in the discrete case in Eq.~(\ref{eq:const_lmax_discrete}). 

We conclude this section by the study of the full probability distribution of $\ell^{\rm \alpha}_{\max,B}(n)$. We perform here the analysis on the exponential case, but it can also be done on the discrete case, leading to the same result in the large $n$ limit. For the exponential case, one obtains the GF of $F^{\e}(\ell,n)_{(0)} = \Pr[\ell^\e_{\max,B}(n) \leq \ell]_{(0)}$ from Eqs. (\ref{eq:def_num_lmax_exp}) and (\ref{eq:def_joint_pdf_exp_partial_int}) as
\beq\label{eq:gf_F}
\sum_{n=0}^\infty F^{\e}(\ell,n)_{(0)} z^n= \sum_{a=1}^\ell z^a \sum_{m=1}^\infty \left[ \tilde h^{\rm e}(z,\ell)\right]^{m-1} q^{\rm e}(m,a) \;.
\eeq
Again, to investigate the large $n$ behavior of $F^{\e}(\ell,n)_{(0)}$, we need to analyze this GF in the limit $z=\e^{-s} \to 1$, i.e., $s\to 0$. From the analysis performed before, using in particular the asymptotic behaviors in Eqs. (\ref{eq:scaling_h_exp}) and (\ref{eq:qe_asympt}) one obtains 
\beq\label{eq:scalingFe}
\sum_{n=0}^\infty F^{\e}(\ell,n)_{(0)} \e^{-s\,n} = \frac{1}{b\,\sqrt{s}} I(s \, \ell) \;,
\eeq
where the function $I(t)$ is given in Eq. (\ref{eq:def_I}). From Eq. (\ref{eq:scalingFe}), we obtain that, for large $n$, $F^{\e}(\ell,n)$ is a function of the ratio $\ell/n$, as expected from our previous computation showing that $\langle \ell^\e_{\max,B}(n)\rangle \sim \lambda_{\max,B}^\e \, n$ (\ref{eq:const_lmax_exp}). This expresses the fact that, for large $n$, the random variable ${\cal R}(n) = {\ell^\e_{\max,B}(n)}/{n}$ becomes independent of $n$, ${\cal R}(n) \to {\cal R}$ when $n \to \infty$, as it is also the case for free RWs \cite{GMS2015}. The cumulative distribution of this random variable $R$, $F_R^{\e}(r)$ is given by
\beq\label{eq:scaling_Fe}
F^{\e}(\ell,n) = F_{\cal R}^{\e}\left(\frac{\ell}{n}=r \right) \;, \; F_{\cal R}^{\e}(r)= 2 \sqrt{\pi} \int_{\Gamma} \frac{\rmd u}{2\pi i} \, \e^{u} \, \frac{I(u\,r)}{\sqrt{u}} \;,
\eeq 
where $\Gamma$ is a Bromwich contour in the complex plane and $r \in [0,1]$. From this expression (\ref{eq:scaling_Fe}), it is possible to study various features of the PDF of $R$, $f^{\e}_{\cal R}(r) = \rmd F_{\cal R}^{\e}(r)/\rmd r$. In particular, one finds that $f^{\e}_{\cal R}(r)$ is non-analytic at $r=1/2$ and one can compute it explicitly on the interval $[1/2,1]$ where it takes a simple form
\beq\label{eq:explicitfR}
f^{\e}_{\cal R}(r) = 1 + \frac{1}{4r^{3/2}} \;, \; \frac{1}{2} \leq r \leq 1 \;.
\eeq
On the other hand, for small $r$, it has an essential singularity where it behaves as 
\beq\label{eq:essential}
\ln f^{\e}_{\cal R}(r) \sim \frac{\ln r}{2r} \;, \; {\rm as} \; r \to 0 \;. 
\eeq
%{\bf Claude: I let you discuss the numerical evaluation and comparison with free RWs.}
The study of the corresponding PDF for free RW was recently carried out in Ref.~\cite{GMS2015} (see also Ref.~\cite{Lamp61}). In this case, the PDF also exhibits a non-analyticity at $r=1/2$ but in this case, it has a square root singularity when $r \to 1$, while for a bridge, there is no singularity [see Eq. (\ref{eq:explicitfR})]. On the other hand, the PDF for free RW also exhibits an essential singularity when $r \to 0$ as in Eq. (\ref{eq:essential}) {\it but} without the logarithmic correction, which is thus a special feature of the RW bridge. In the next section, we present a numerical evaluation of this PDF $f^{\e}_{\cal R}(r)$ (see the left panel of Fig. \ref{fig:R}).  

\section{Numerical results}\label{sec:simu}

In this section, we present results obtained from numerical simulations of random walk bridges, which we compare to our exact analytical computations. We consider discrete ($\alpha = {\rm d}$) as well as continuous ($\alpha = {\rm c}$) jump distributions, for $\mu =2$, with a special attention on the exponential jump distribution ($\alpha = {\rm e}$).

\subsection{Discrete random walk bridges}

\begin{figure}[!ht]
\begin{center}
\includegraphics[angle=0,width=.8\linewidth]{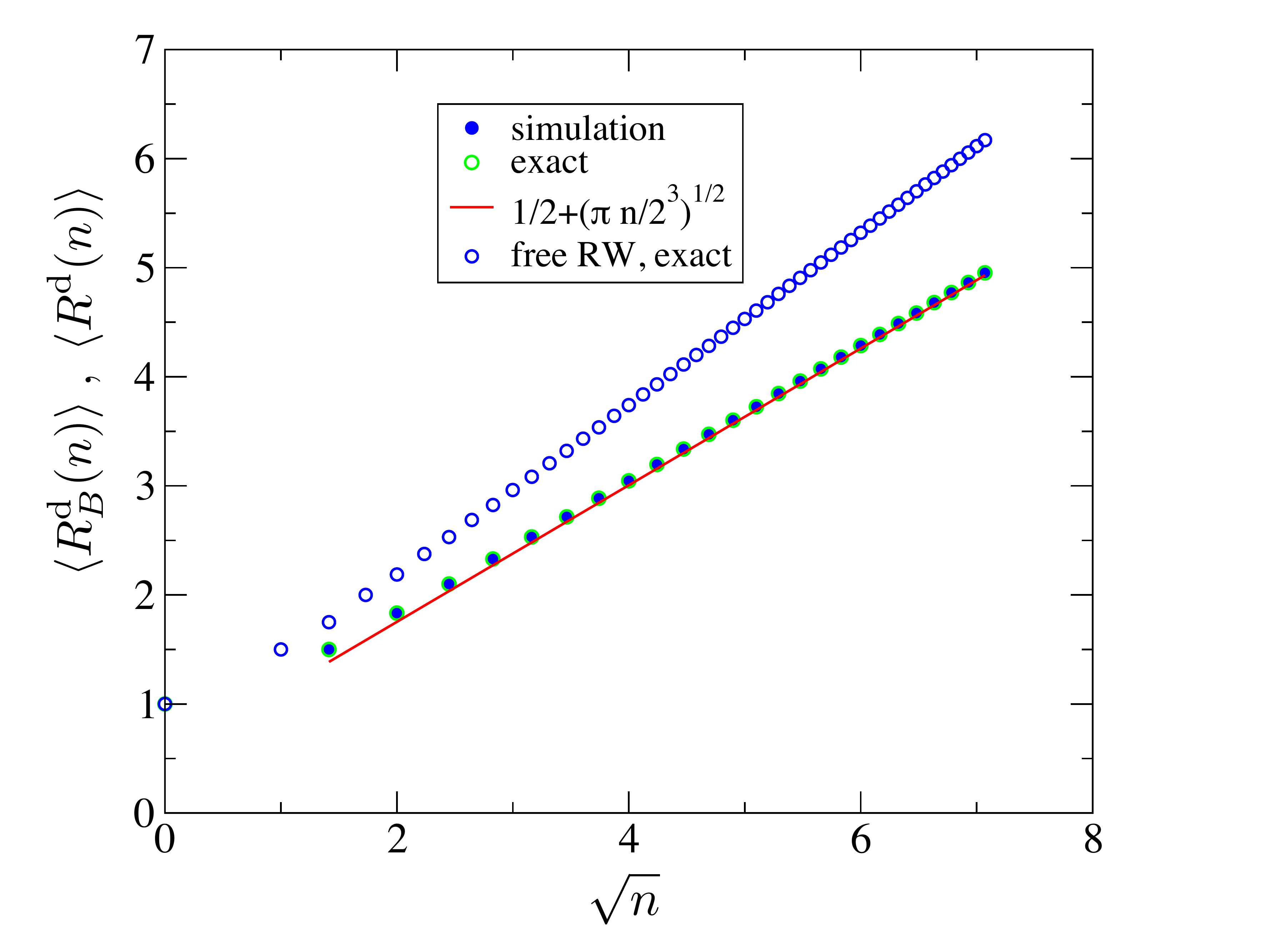}
\caption{
Average number of records $\langle R^{\rm d}_B(n)\rangle$ as a function of $\sqrt{n}$ for discrete RW bridges. The full blue circles are the results of numerical simulations. The green empty circles correspond to the exact results given in Eq. (\ref{eq:exact_av_Rn_discrete}), while the full red line is a plot of its asymptotic form given in Eq. (\ref{eq:av_rec_first_correction}). For comparison, we have also plotted, with empty blue circles the exact results for $\langle R^{\rm d}(n)\rangle$ for the free RW, as obtained in \cite{MZ2008}, which displays a faster growth $\langle R^{\rm d}(n)\rangle \sim (2/\sqrt{\pi}) \sqrt{n}$ for $n \gg 1$.}
\label{fig:RnDiscrete}
\end{center}
\end{figure}

To simulate discrete RW bridges, we generate simple free random walks of $n$ steps according to Eqs. (\ref{def_RW}) and (\ref{def_discrete}) and retain only the walks such that $x(n)=x(0)=0$. The statistics is then performed over this restricted ensemble of RW bridges. In Fig.~\ref{fig:RnDiscrete}, we show a plot of the average number of records $\langle R_B^{\rm d}(n)\rangle$, computed numerically, as a function of $\sqrt{n}$, which shows a perfect agreement with our exact results in (\ref{eq:exact_av_Rn_discrete}). In Fig.~\ref{fig:RnDiscrete} we have also plotted the asymptotic estimate of $\langle R_B^{\rm d}(n)\rangle$ beyond the leading ${\cal O}(\sqrt{n})$ term. Indeed, from the exact formula in Eq. (\ref{eq:exact_av_Rn_discrete}), one obtains:
\begin{eqnarray}\label{eq:av_rec_first_correction}
\langle R_B^{\rm d}(n)\rangle \sim \frac{\sqrt{\pi}}{2^{3/2}} \sqrt{n} + \frac{1}{2} + o(1) \;,
\end{eqnarray} 
which, as can be seen on Fig. \ref{fig:RnDiscrete}, is a very good estimate of the exact result for $\sqrt{n} \geq 5$.  

\begin{figure}[!ht]
\begin{center}
\includegraphics[angle=0,width=.8\linewidth]{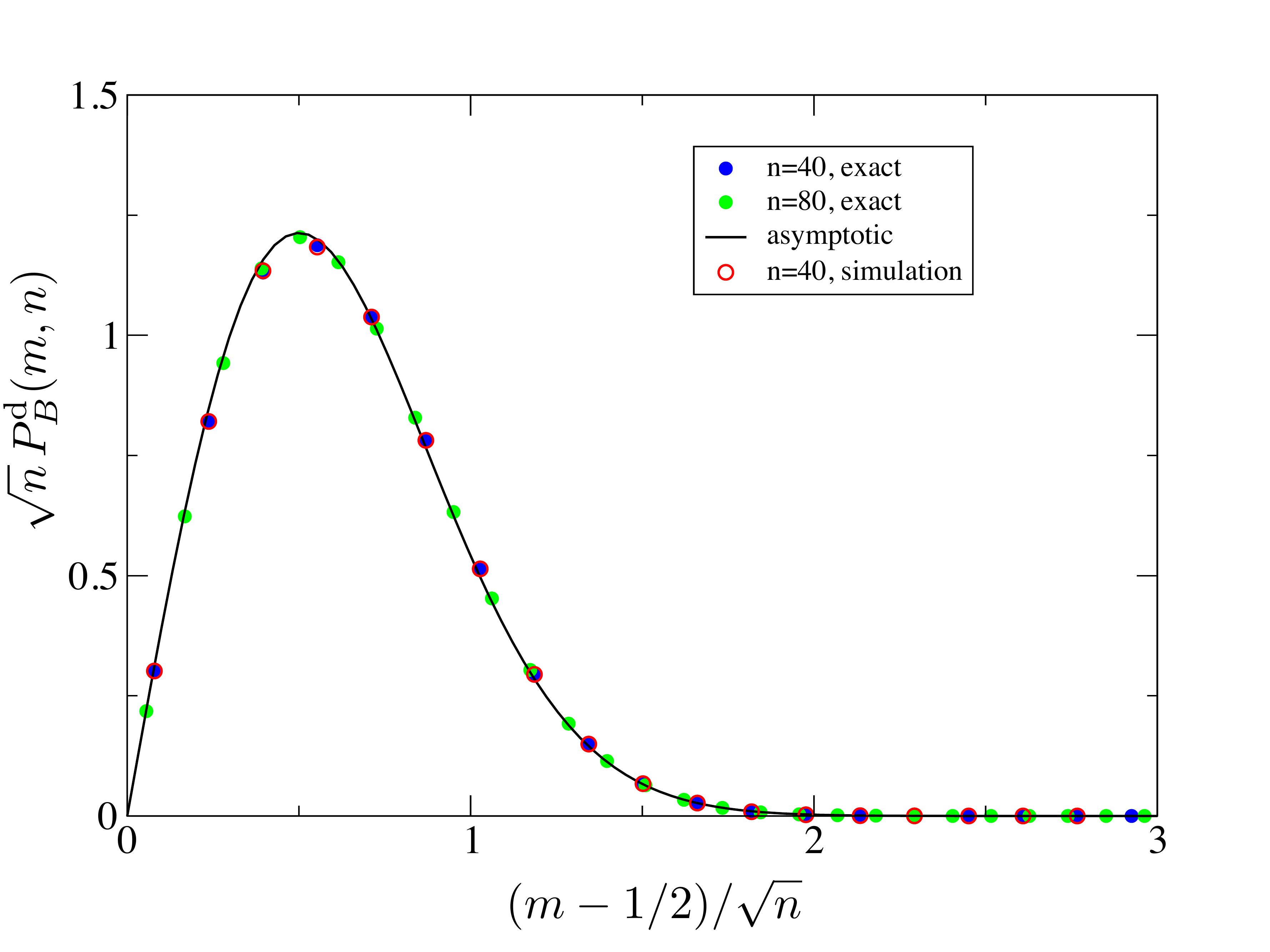}
\caption{Plot of $\sqrt{n}\,P_B^{\rm d}(m,n)$ as a function of $(m-1/2)/\sqrt{n}$ for discrete RW -- see Eq. (\ref{eq:correction_scaling}) and the discussion below in the text. The filled circles correspond to the exact values obtained from Eq. (\ref{eq:explicit_dist_rn_discrete}) for different values of $n = 40$ (blue) and $n=80$ (green). The open circles correspond to the results of numerical simulations for $n=40$, which coincide exactly with the exact results. Finally, the solid line is the exact scaling function $\varphi_B^{\rm d}(X)$ given in Eq. (\ref{eq:summary_av_R_discrete}).  
}
\label{fig:fRnDiscrete}
\end{center}
\end{figure}

We have also computed numerically the distribution of $R_B^{\rm d}(n)$, $P_B^{\rm d}(m,n) = \Pr(R_B^{\rm d}(n)=m)$ both exactly, for $n=40$ and $80$ from the analytical formula in Eq.~(\ref{eq:explicit_dist_rn_discrete}) and numerically for $n=40$. As shown in Fig. \ref{fig:fRnDiscrete}, the numerical results agree perfectly with our exact analytical results.   
Besides, for large $n$ and large $m$, keeping $m/\sqrt{n}=X$ fixed, one expects the scaling form given in Eq. (\ref{eq:summary_pdf_discrete}). Our simulations for finite values of $n$ exhibit small finite size corrections to this limiting scaling form. As suggested in Fig. \ref{fig:fRnDiscrete}, these finite $n$ actually correspond to a simple shift of the scaling variable $m/\sqrt{n} \to (m-1/2)/\sqrt{n}$, i.e., 
\begin{eqnarray}\label{eq:correction_scaling}
P_B^{\rm d}(m,n) \sim \frac{1}{\sqrt{n}} \varphi_B^{\rm d}\left(\frac{m-1/2}{\sqrt{n}} \right) \;,
\end{eqnarray}
where the scaling function $\varphi_B^{\rm d}(X)$ is given in Eq. (\ref{eq:summary_pdf_discrete}). Although we have not performed a detailed analytical study of the finite $n$ corrections to the limiting PDF in Eq. (\ref{eq:summary_pdf_discrete}), this form in Eq. (\ref{eq:correction_scaling}), and in particular this shift of $1/2$, is consistent with the asymptotic expansion of the average number $\langle R_B^{\rm d}(n)\rangle$, beyond the leading ${\cal O}(\sqrt{n})$ term, given in Eq.~(\ref{eq:correction_scaling}). 

\begin{figure}[!ht]
\begin{center}
\includegraphics[width=0.8\linewidth]{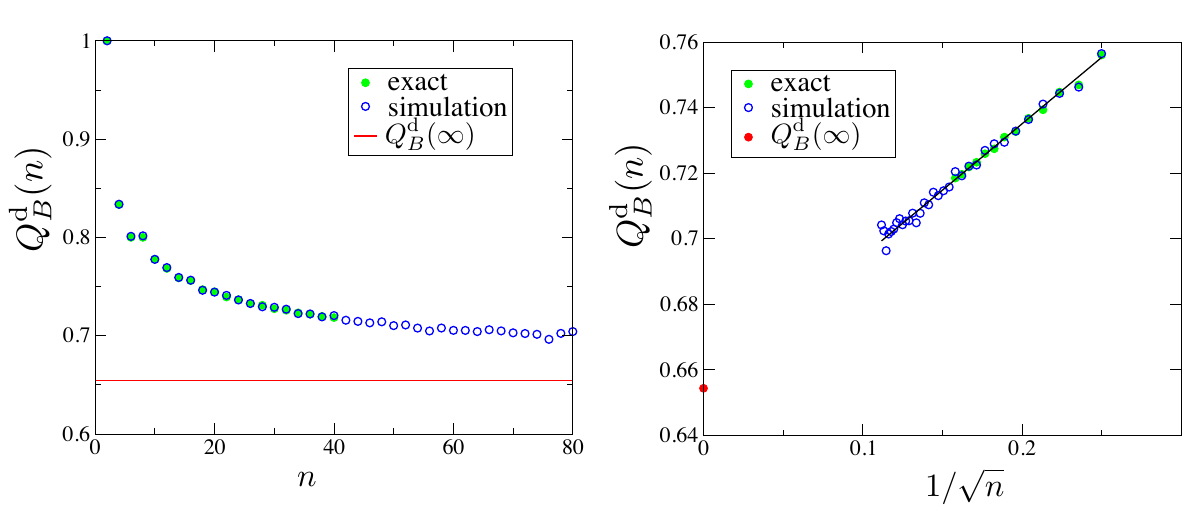}
\caption{{\bf Left:} Probability of record breaking $Q_B^{\rm d}(n)$ for the discrete random walk bridge. The blue open circles correspond to 
our numerical simulations while the yellow full circles are the exact values of $Q_B^{\rm d}(n)$, extracted from the generating function in Eq. (\ref{eq:num_Q_discrete}). {\bf Right:} Slow convergence of $Q_B^{\rm d}(n)$ to its asymptotic value $Q_B^{\rm d}(\infty) = 0.6543037 \ldots$ in Eq. (\ref{eq:cQ}). The full line is a guide to the eyes and corresponds to $Q_B^{\rm d}(\infty) + {\rm const.}/\sqrt{n}$.}
\label{fig:Qn}
\end{center}
\end{figure}

Then, we have computed numerically the probability of record breaking $Q_B^{\rm d}(n)$, both from numerical simulations and from 
our exact formula in Eq. (\ref{eq:num_Q_discrete}). As shown in the left panel of Fig. \ref{fig:Qn}, the agreement between both estimates is very good. Besides, we also see on that figure that the convergence of $Q_B^{\rm d}(n)$ to its asymptotic value $Q_B^{\rm d}(\infty)$ is quite slow. The right panel of Fig. \ref{fig:Qn} indicates that the speed of convergence is proportional to $1/\sqrt{n}$, i.e., $Q_B^{\rm d}(n) - Q_B^{\rm d}(\infty) \propto 1/\sqrt{n}$. Furthermore, the estimate of $Q_B^{\rm d}(\infty)$ obtained in this way, by extrapolation, is in good agreement with our exact results in Eq. (\ref{eq:cQ}), see the right panel of Fig. \ref{fig:Qn}.
% figure

\begin{figure}[!ht]
\begin{center}
\includegraphics[width=0.8\linewidth]{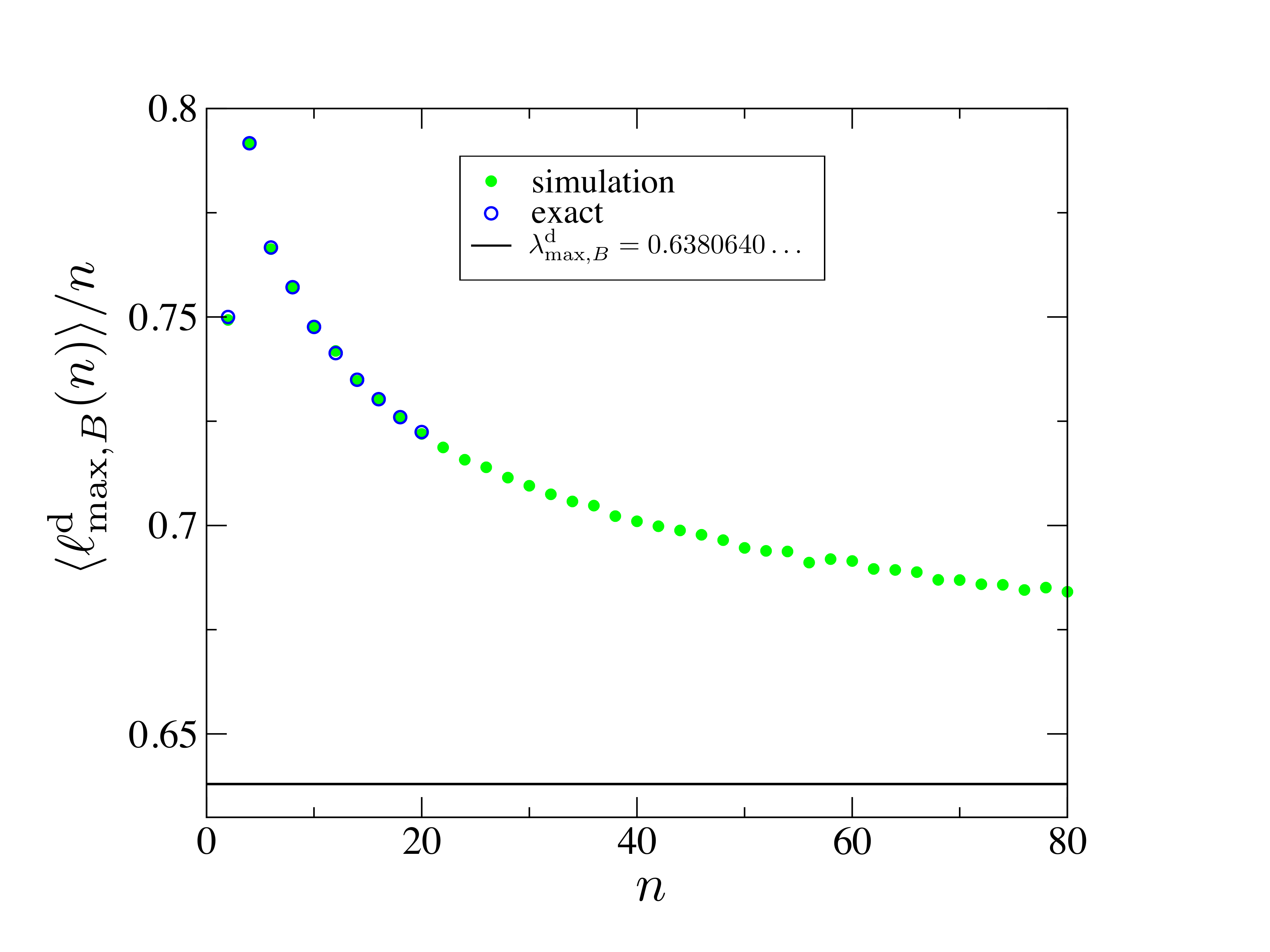}
\caption{Plot of $\langle \ell_{\max,B}^{\rm d}(n)\rangle/n$ as a function of $n$ for the discrete random walk bridge. The open circles correspond to the exact analytical value computed from the generating function in Eq.~(\ref{eq:expr_lmax_2}) while the full circles correspond to the estimates obtained from direct numerical simulations. The solid line corresponds to the asymptotic value $\lambda_{\max,B}^{\rm d} = 0.6380640\ldots$, see Eq. (\ref{eq:intro_const_lmax_exp}).}\label{fig:lmax_dis}
\end{center}
\end{figure}

Finally, in Fig. \ref{fig:lmax_dis}, we show numerical results for $\langle \ell_{\max,B}^{\rm d}(n)\rangle/n$, obtained from numerical simulations and compare them to the exact results which can be straightforwardly obtained from the generating function in Eq. (\ref{eq:expr_lmax_2}). Here also the agreement between numerics and theory is very good. We also notice that, as in the case of the probability of record breaking (see Fig. \ref{fig:Qn}), the convergence of $\langle \ell_{\max,B}^{\rm d}(n)\rangle/n$ to the asymptotic value $\lambda_{\max,B}^{\rm d}$ [see Eq. (\ref{eq:intro_const_lmax_exp})] is actually quite slow.

\subsection{Continuous jump distribution with $\mu = 2$.}

To generate a RW bridge with continuous jump distributions and $\mu = 2$ (i.e., with a finite variance $\sigma$), we first generate a free random walk of $n$ steps, starting from $x(0)=0$ and evolving according to Eq. (\ref{def_RW}). We subsequently use the following construction
\begin{eqnarray}\label{def_construction_bridge}
x_B(k) = x(k) - \frac{k}{n} x(n) 
\end{eqnarray}
to generate a RW bridge $\{x_B(k)\}$, for $0 \leq k \leq n$. In particular, Eq. (\ref{def_construction_bridge}) implies obviously $x_B(0) = x_B(n) = 0$. This simple construction (\ref{def_construction_bridge}) holds, for any finite $n$, for Gaussian jump distributions. For other jump distributions, including in particular the exponential jump distribution ($\alpha = {\rm e}$), this construction is only approximate, becoming exact in the limit of large $n$, where $x_B(k)/\sqrt{n}$, for $k \sim {\cal O}(n)$, converges to the Brownian bridge. 
\begin{figure}[!ht]
\begin{center}
\includegraphics[width=.8\linewidth]{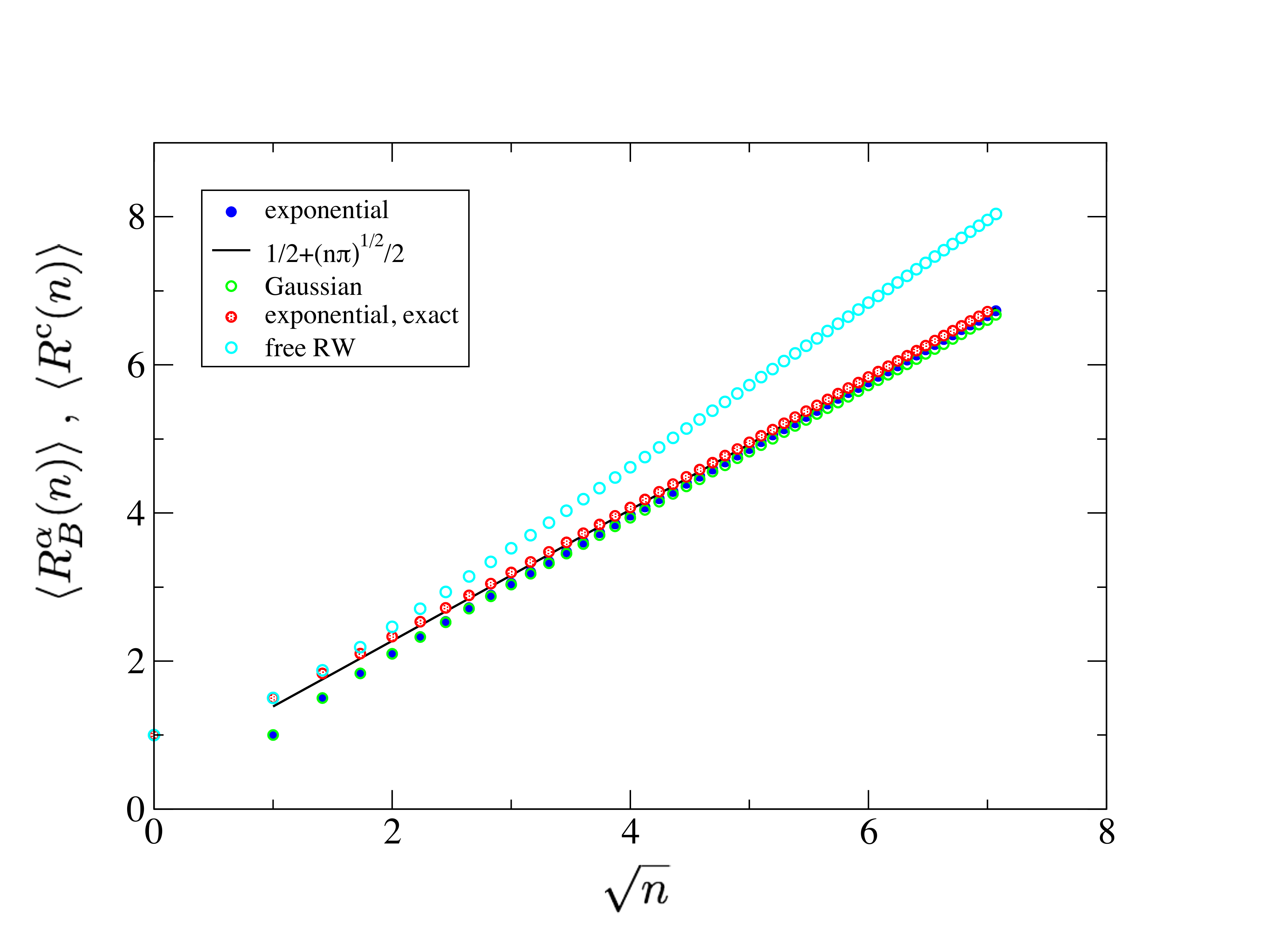}
\caption{
Average number of records $\langle R^{\alpha}_B(n)\rangle$ obtained by simulations of a RW with an exponential jump distribution, $\alpha = {\rm e}$,   (blue circles) and for a Gaussian jump distribution, $\alpha = {\rm c}$, (green circles), compared to the exact asymptotic value (full black line), given in Eq. (\ref{eq:corrections_exp}). The red circles correspond to the result for the exponential case in Eq. (\ref{eq:exact_av_Rn_exp}). The slight discrepancy between the exact and the numerical results in the exponential case for small values of $n$ is due to the fact that our numerical procedure to generate a bridge (\ref{def_construction_bridge}) is only approximate for finite $n$. For comparison, we have also plotted (with cyan circles) the exact results for a free RW with a continuous jump distribution obtained in \cite{MZ2008}, which displays a faster growth $\langle R^{\rm c}(n)\rangle \sim 2/\sqrt{\pi}\sqrt{n}$ for $n \gg 1$.
}
\label{fig:Rn_C}
\end{center}
\end{figure}
In Fig. \ref{fig:Rn_C}, we show a plot of the average number of records $\langle R^{\alpha}_B(n)\rangle$ both for an exponential distribution, $\alpha = {\rm e}$, and for a Gaussian distribution, both of them belonging to the class of continuous jump distributions $\alpha = {\rm c}$ as in Eq. (\ref{def_p_c}) with $\mu = 2$. For large $n$, one expects that, in both cases, $\langle R^{\rm e}_B(n)\rangle \sim \langle R^{\rm c}_B(n)\rangle \sim A_B^{\rm c}(\mu=2)\sqrt{n}$ with $A_B^{\rm c}(\mu=2) = \sqrt{\pi}/2$ [see Eqs. (\ref{eq:av_records_bridge}) and (\ref{eq:av_records_ampli_bridge})], in good agreement with our numerical simulations (see Fig. \ref{fig:Rn_C}). In fact, our numerical simulations suggest that the first correction to this asymptotic behavior is the same for the exponential and the Gaussian cases and can be computed from our exact formula for $\langle R^{\rm e}_B(n)\rangle$ in Eq. (\ref{eq:exact_av_Rn_exp}), yielding
\begin{eqnarray}\label{eq:corrections_exp}
\langle R_B^{\rm e}(n)\rangle \sim \frac{\sqrt{\pi}}{2}\sqrt{n} + \frac{1}{2} + o(1) \;.
\end{eqnarray}
Note that the first ${\cal O}(1)$ correction, namely $1/2$, is the same as in the discrete case (\ref{eq:av_rec_first_correction}). For smaller values of $n$, the data for the exponential case, $\alpha = {\rm e}$, in Fig. \ref{fig:Rn_C} show a slight discrepancy between the numerical results and our exact formula for $\langle R_B^{\e}(n)\rangle$ given in Eq. (\ref{eq:exact_av_Rn_exp}). This is due to the fact that our construction of the RW walk bridge (\ref{def_construction_bridge}) is in this case only approximate for small values of $n$. Note that to simulate exact bridges in this exponential case, one could use the Monte-Carlo method used in Ref. \cite{SM2010}, which however requires more numerical efforts and goes beyond the scope of the present work. 

\begin{figure}[!ht]
\begin{center}
\includegraphics[width=0.8\linewidth]{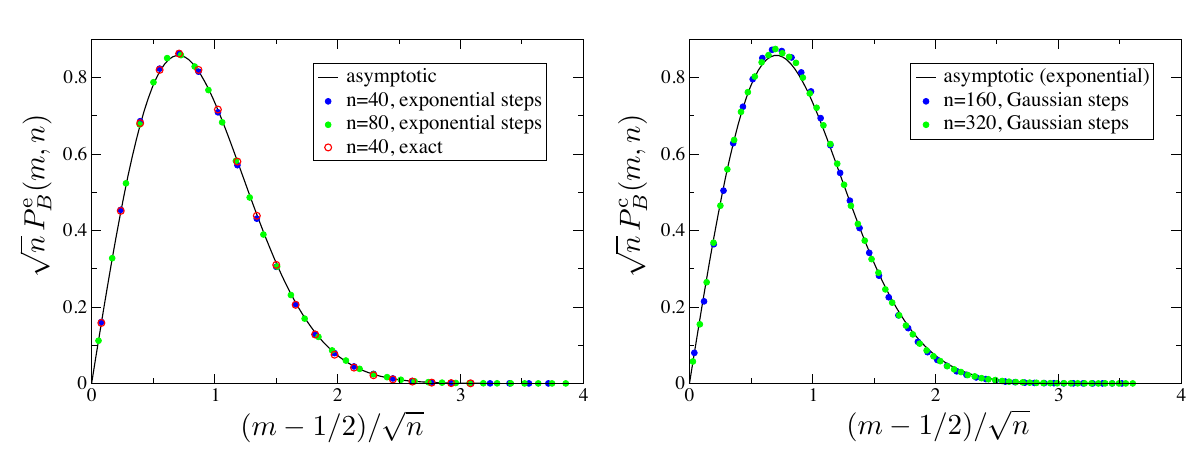}
\caption{{\bf Left:} Plot of $\sqrt{n}\,P_B^{\rm e}(m,n)$ as a function of $(m-1/2)/\sqrt{n}$ for an exponential jump distribution. The filled circles correspond to simulations for $n=40$ (blue circles) and $n=80$ (green circles). The empty circles correspond to the exact analytical result given in Eq. (\ref{eq:dist_rn_explicit}) for $n=40$. The solid line corresponds to the exact scaling function $\varphi_B^{\rm e}(X)$ given in Eq. (\ref{eq:summary_pdf_exp}). {\bf Right:} Plot of $\sqrt{n}\,P_B^{\rm c}(m,n)$ as a function of $(m-1/2)/\sqrt{n}$ for a Gaussian jump distribution for $n= 160$ and $n=320$. The solid line corresponds to the exact scaling function $\varphi_B^{\rm e}(X)$ given in Eq. (\ref{eq:summary_pdf_exp})
}
\label{fig:PRn_C}
\end{center}
\end{figure}

In both cases, exponential and Gaussian jump distributions, we have computed the distribution of the number of records $P_B^{\rm e}(m,n)$ and $P_B^{\rm c}(m,n)$. The results are shown in Fig. \ref{fig:PRn_C}. 
In the left panel, we show the results for an exponential jump distribution, $\alpha = \e$, which show a very good agreement between our numerical estimates and the exact formula for $P_B^{\rm e}(m,n)$ given in Eq. (\ref{eq:dist_rn_explicit}). These data for finite $n$ show some deviation from the asymptotic result in Eq. (\ref{eq:summary_pdf_exp}) valid for $n \gg 1$. As suggested in Fig.~\ref{fig:fRnDiscrete}, and similarly to the discrete case (\ref{eq:correction_scaling}), these finite $n$ corrections actually correspond to a simple shift of the scaling variable $m/\sqrt{n} \to (m-1/2)/\sqrt{n}$, i.e., 
\begin{eqnarray}\label{eq:correction_scaling_exp}
P_B^{\rm e}(m,n) \sim \frac{1}{\sqrt{n}} \varphi_B^{\rm e}\left(\frac{m-1/2}{\sqrt{n}} \right) \;,
\end{eqnarray}
where the scaling function $\varphi_B^{\rm e}(X)$ is correctly predicted by the expression given in Eq.~(\ref{eq:summary_pdf_exp}). Note that this shift of $1/2$ is fully consistent with the asymptotic expansion of $\langle R_B^{\rm e}(n)\rangle$ beyond the leading order in Eq. (\ref{eq:corrections_exp}). In the right panel of Fig. \ref{fig:PRn_C}, we show numerical data for $P_B^{\rm c}(m,n)$ for a Gaussian jump distribution. These data are particularly instructive, as we do not have any analytical result for this quantity for Gaussian jumps. Our data actually suggest that the distribution of the number of records in this case is also described, for large $n$, by the same scaling form as for the exponential jump distribution in Eq. (\ref{eq:correction_scaling_exp}), with the same scaling function $\varphi_B^\e(X)$. These data support our conjecture that this result in Eq. (\ref{eq:summary_pdf_exp}), which we can explicitly show only for the exponential case, should actually hold for any continuous jump distribution as in Eq. (\ref{def_p_c}) with $\mu = 2$.  

\begin{figure}[!ht]
\begin{center}
\includegraphics[angle=0,width=.8\linewidth]{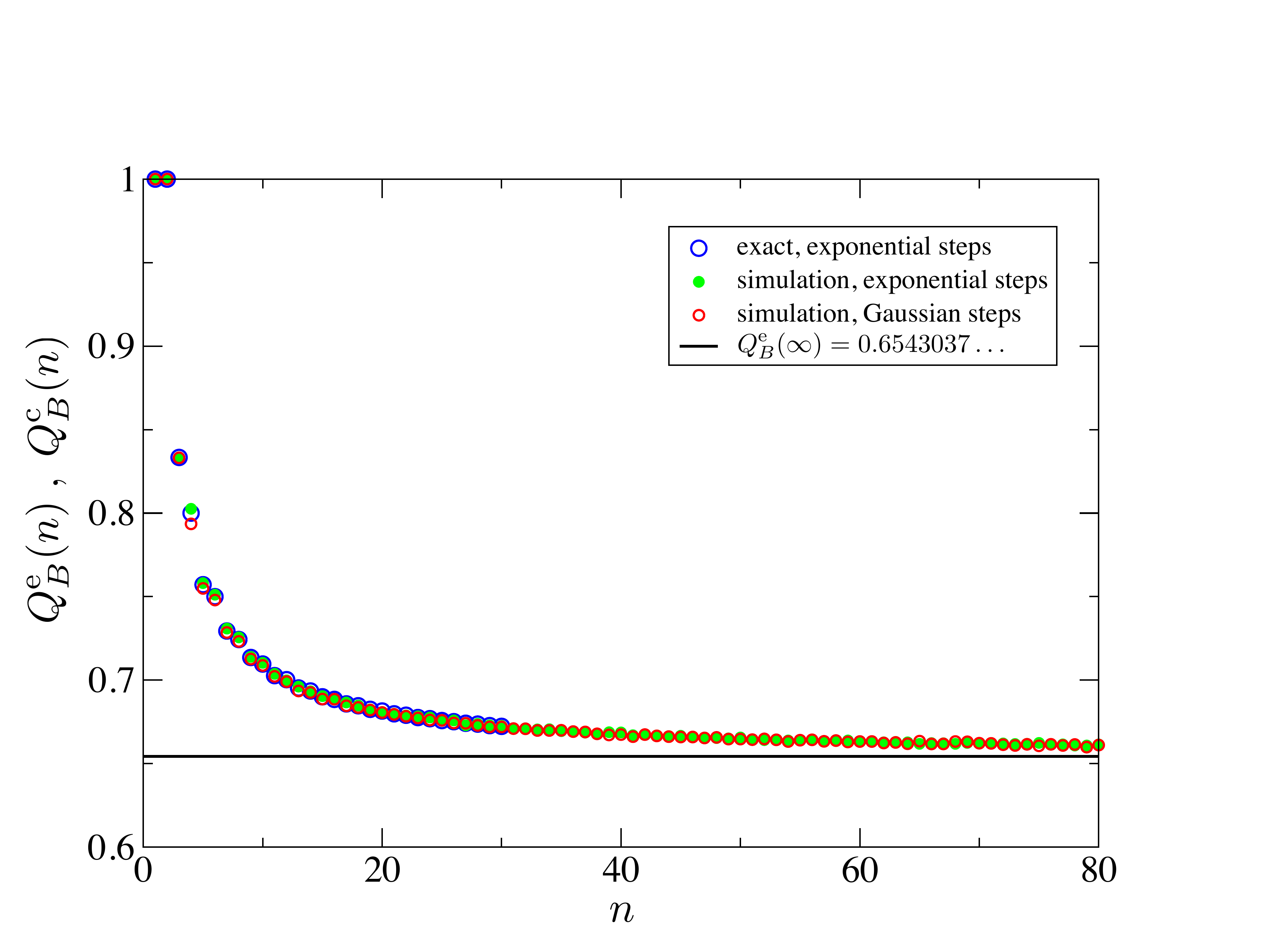}
\caption{Probability of record breaking for an exponential jump distribution, $Q^{\e}_B(n)$, and for a Gaussian jump distribution, $Q^{\rm c}_B(n)$. In the exponential case, we show data corresponding to the exact result (open circle), obtained from the generating function in Eq. (\ref{eq:num_Q_exp}) and data obtained from direct numerical simulations (full circles) using Eq. (\ref{def_construction_bridge}). The data for a Gaussian jump distribution correspond to numerical simulations using Eq. (\ref{def_construction_bridge}). The convergence towards the exact asymptotic value $Q_B^{\e}(\infty) = 0.654304\ldots$ is fast ($\propto 1/n$).}
\label{fig:Q}
\end{center}
\end{figure}

We have also computed numerically the probability of record breaking $Q^{\e}_B(n)$ for an exponential jump distribution, which we have plotted in Fig. \ref{fig:Q}. Here again, we see a very good agreement between our exact results [extracted from the generating function in Eq.~(\ref{eq:num_Q_exp})] and the numerical estimates. These data also corroborate the asymptotic behavior $Q_B^\e(n) \to Q_B^\e(\infty) = 0.654304 \ldots$ when $n \to \infty$, where $Q_B^\e(\infty) = Q_B^{\rm d}(\infty)$ is given in Eq. (\ref{eq:cQ}) and coincides with the asymptotic value $Q_B^{\rm d}(\infty)$ found for the discrete RW. Note that our numerical results indicate that the convergence to the asymptotic value is $\propto 1/n$, which is thus faster than for the discrete case, where the convergence is $\propto 1/\sqrt{n}$ (see the right panel of Fig. \ref{fig:Qn}). In Fig. \ref{fig:Q}, we have also plotted the estimates of $Q^{\rm c}_B(n)$ for random walk bridges, with Gaussian jump distribution, obtained from numerical simulations, using Eq. (\ref{def_construction_bridge}). The data for exponential and Gaussian jumps are, for $n$ sufficiently large, indistinguishable, which reinforce our claim that all our asymptotic results for the exponential case should also hold for any continuous jump distributions with $\mu = 2$, in the limit $n \to \infty$. 

\begin{figure}[!hh]
\begin{center}
\includegraphics[angle=0,width=.8\linewidth]{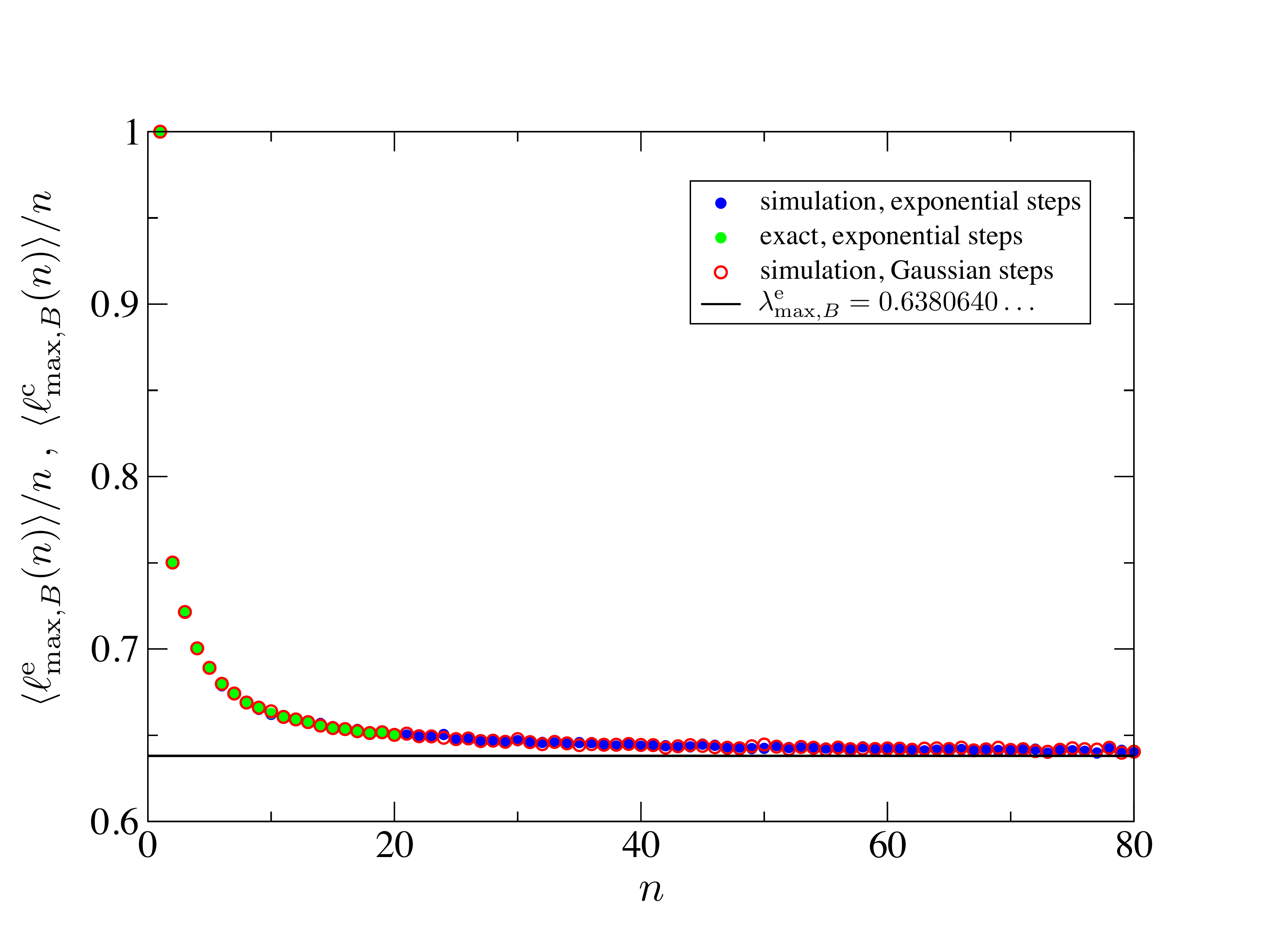}
\caption{Plot of $\langle \ell^{\rm e}_{\max,B}(n)\rangle/n$, for an exponential jump distribution, and $\langle \ell^{\rm c}_{\max,B}(n)\rangle/n$, for a Gaussian jump distribution, as a function of $n$. For the exponential case, the exact results are obtained from Eq. (\ref{eq:def_num_lmax_exp}), while in both cases, the numerical estimates are obtained by simulating the random walk bridges from Eq. (\ref{def_construction_bridge}), with the appropriate jump distribution.}
\label{fig:lmax_c}
\end{center}
\end{figure}

Similarly, in Fig. \ref{fig:lmax_c}, we show a plot of $\langle \ell^{\rm e}_{\max,B}(n)\rangle/n$, for an exponential jump distribution and of $\langle \ell^{\rm c}_{\max,B}(n)\rangle/n$, for a Gaussian jump distribution. In the exponential case, we have plotted both our exact results, which can be derived from Eq. (\ref{eq:def_num_lmax_exp}), and the estimates obtained by simulating the random walk bridge, using Eq. (\ref{def_construction_bridge}). We see that the agreement is very good. Interestingly, we also see that the data for exponential and Gaussian jump distributions are almost indistinguishable for $n$  sufficiently large.   

\begin{figure}[!hh]
\begin{center}
\includegraphics[width = 0.8\linewidth]{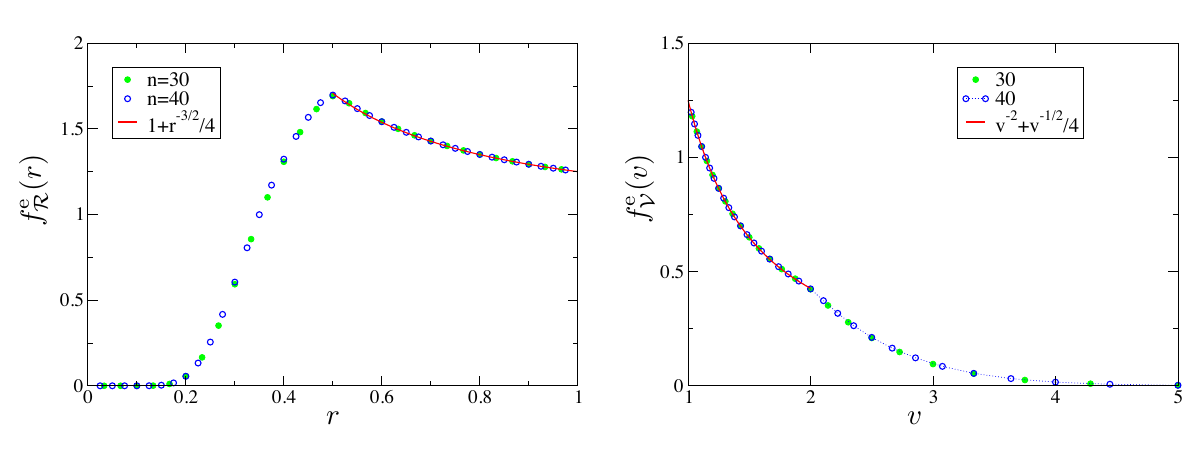}
\caption{{\bf Left:} Probability density of the ratio ${\cal R}(n)=\l_{\max}(n)/n$ for an exponential jump distribution, for $n=30$ and $n=40$.
The points were obtained from the exact expression for the generating function of the cumulative distribution $F^\e(\ell,n)$ (\ref{eq:gf_F}). 
The red curve is the analytical prediction for the limiting PDF $f_R^\e(r)$~(\ref{eq:explicitfR}) valid for $1/2\leq r \leq 1$. {\bf Right:} Probability density of ${\cal V}(n)=n/\l_{\max}(n)$ for an exponential jump distribution of steps, for $n=30$ and $n=40$, 
obtained from the left panel.}
\label{fig:R}
\end{center}
\end{figure}

Finally, we present the numerical evaluation of the distribution $f_{\cal R}^\e(n) = \rmd F_{\cal R}^{\e}(r)/\rmd r$ of ${\ell_{\max,B}^\e}(n)/n$ in the large $n$ limit [see Eq. (\ref{eq:scaling_Fe})]. The numerical estimates of $f_{\cal R}^\e(r)$ can be obtained by evaluating $F^\e(\ell,n)$ from the generating function that was explicitly computed in Eq.~(\ref{eq:gf_F}), which we expand close to $z=0$. The same procedure was used by us in Ref. \cite{GMS2015} to plot the corresponding PDF for the free RW. The plot of $f_{\cal R}^\e(r)$, for RW bridges with an exponential jump distribution, obtained this way is shown in the left panel of Fig. \ref{fig:R}. The non-analyticity at $r=1/2$ is clearly visible and in addition, we observe a perfect agreement between the numerical evaluation of $f_{\cal R}^\e(r)$ and the exact result for $1/2 \leq r \leq 1$ [see Eq. (\ref{eq:explicitfR})]. Finally, in the right panel of Fig. \ref{fig:R}, we show, for completeness, a plot of the PDF of ${\cal V}= 1/{\cal R}$, whose PDF turns out to be simpler to study (see Ref. \cite{GMS2015} as well as \ref{sec:appendix_PDF}).  

We conclude this section by noticing that a natural way of simulating numerically random walk bridges would be to write an effective local
equation of motion for $x_B(k)$ which would take into account the conditioning that the walker has to return back to the origin after $n$ time
steps. For Brownian motion, which is continuous both in space and time, this can indeed be done by writing an effective Langevin equation
for the Brownian bridge \cite{Doo1957, MO2015}. Extending this approach to discrete time random walks -- both with discrete and continuous jumps -- is a very interesting open problem.

\section{Conclusion}\label{sec:conclusion}

To conclude, we have studied the record statistics of RW bridges, for different types of symmetric jump distributions. Our results show that the record statistics of such constrained random walks, which are constrained to start and end at the origin after $n$ steps, are quantitatively different from their counterpart for free RWs, which can end up anywhere on the real axis. We first showed that the statistics of the number of records $R^{\alpha}_B(n)$ is not only different for discrete ($\alpha = \rmd$) and continuous ($\alpha = {\rm c}$) distributions but, even for continuous jump distributions, also depends on the details of this distribution. We obtained exact results for the average number of records $\langle R_B^\alpha(n)\rangle \sim A_B^\alpha \sqrt{n}$ and showed in particular that $A_B^{\rm c}\equiv A_B^{\rm c}(\mu)$ depends continuously on the L\'evy index $\mu$ characterizing the RW. This is quite different from the free RW case where $R^{\rm c}(n) \sim A^{\rm c}\sqrt{n}$ where $A^{\rm c} = 2/\sqrt{\pi}$, independently of $\mu$. Furthermore, we have computed exactly the full record statistics for two different types of jump distributions: the discrete RW ($\alpha = {\rm d}$) and the exponential distribution ($\alpha = \e$). We emphasize that these calculations are technically much harder than for free RWs as the records statistics of RW bridges requires to keep track not only of the ages of the records but also of the increments between successive records [see Fig. \ref{Fig1} and Eq. (\ref{eq:def_joint_pdf_exp_2})]. For these two jump distributions, we have also computed the probability of record breaking $Q_{B}^{\alpha}(n)$ and the average age of the longest lasting record $\langle \ell^\alpha_{\max,B}(n) \rangle$ and shown that they give rise to two new non-trivial constants which we have computed explicitly (see Table 1). 

Although the records statistics for a continuous jump distribution depends, for finite $n$, on the details of this distribution, one expects that, for large $n$, it only depends on the L\'evy index $\mu$ in Eq.~(\ref{def_p_c}). Although we can not prove this statement, our numerical data indicate that this should indeed be the case, at least for $\mu = 2$. This implies that our asymptotic results (see Table \ref{table:summary}) obtained in the exponential case, which we could solve exactly, should describe the large $n$ limit of any continuous jump distribution with $\mu = 2$. The generalization of our results to arbitrary value of the L\'evy index $0 < \mu<2$ remains a challenging open question.

\ack
 SNM and GS acknowledge the Indo-French Centre for the Promotion of Advanced Research under Project 4604-3.

\appendix

\section{Useful formulas for discrete random walks}\label{sec:appendix_discrete}

Here we consider a discrete RW, starting from $x(0) = 0$ and evolving via
\beq\label{def_rw_app}
x(k) = x(k-1) + \eta(k) \;,
\eeq
where the jump variables $\eta(k)$'s are i.i.d. random variables distributed according to $p^{\rmd}(\eta) = \frac{1}{2}\delta(\eta+1) + \frac{1}{2}\delta(\eta-1)$. Let us denote by $W(x,n)$ the number of lattice RW (\ref{def_rw_app}) starting at $x(0)=0$ and ending in $x$ after $n$ time steps. One has
\beq\label{eq:W}
W(x,n) = 
\begin{cases}
&{n \choose \frac{n+x}{2}} \;, \; {\rm if} \; n+x \; {\rm is \; even} \\
&0 \; {\rm if} \; n+x \; {\rm is \; odd} \;.
\end{cases}
\eeq
From $W(x,n)$, one obtains immediately the discrete propagator $G^{\rmd}(x,x_0,n)$ as
\beq\label{eq:G}
G^{\rmd}(x,x_0,n) = \frac{W(x-x_0,n)}{2^n} 
\begin{cases}
&\frac{1}{2^n}{n \choose \frac{n+x-x_0}{2}} \;, \; {\rm if} \; n+x-x_0 \; {\rm is \; even} \\
&0 \; {\rm if} \; n+x-x_0 \; {\rm is \; odd} \;.
\end{cases}
\eeq
The generating function $\tilde G^{\rmd}(x,x_0,n)$ of $G^{\rmd}(x,x_0,n)$ can be computed from this explicit expression (\ref{eq:G}) and it yields:
\beq\label{GF_Gfree_d}
\tilde G^{\rmd}(x,x_0,n) = \sum_{n=0}^\infty G^{\rmd}(x,x_0,n) z^n &=& \sum_{k=\lceil \frac{x-x_0}{2}\rceil}^\infty \left( \frac{z}{2}\right)^{2k-(x-x_0)} {2k-(x-x_0) \choose k} \nonumber \\ 
&=& \frac{1}{\sqrt{1-z^2}} \left(\frac{1-\sqrt{1-z^2}}{z} \right)^{x-x_0} \;,
\eeq
where $\lceil u \rceil$ is the smallest integer not less than $u$ and where the last equality can be obtained using, for instance, Mathematica.

Furthermore, from the expression of $G^{\rmd}(x,x_0,n)$ (\ref{eq:W}), one can also compute the constrained propagator $G^{\rmd}_{\geq}(x,0,n)$ using the method of images. One has indeed:
\beq\label{eq:ggeq_app}
G^{\rmd}_{\geq}(x,0,n) &=& G^{\rmd}(x,0,n) - G^{\rmd}(x,-2,n) \\
&=& 
\begin{cases}
&\frac{1}{2^n} \left({n \choose \frac{n+x}{2}} - {n \choose \frac{n+x}{2}+1} \right) \;, \; {\rm if} \; n+x \; {\rm is \; even} \\
&0 \;, \; {\rm if} \; n+x \; {\rm is \; odd}
\end{cases} \;.
\eeq
The generating function $\tilde G_{\geq}^{\rmd}(x,0,z)$ of $G_{\geq}^{\rmd}(x,0,n)$ can then be obtained from this explicit expression (\ref{eq:ggeq_app}) as
\beq\label{GF_Ggeq_d}
\hspace*{-1.5cm}\tilde G_{\geq}^{\rmd}(x,0,z) &=& \sum_{n=0}^\infty G_{\geq}^{\rmd}(x,0,n) z^n \nonumber \\
&=& \sum_{k=\lceil \frac{x}{2}\rceil}^\infty \left( \frac{z}{2}\right)^{2k-x} {2k-x \choose k} - \sum_{k=\lceil \frac{x+2}{2}\rceil}^\infty \left( \frac{z}{2}\right)^{2k-(x+2)} {2k-(x+2) \choose k} \nonumber \\
&=& \frac{2}{z} \left(\frac{1-\sqrt{1-z^2}}{z} \right)^{x+1} \;.
\eeq
Note that this result (\ref{GF_Ggeq_d}) which we obtained here by counting paths can also be obtained using a backward Fokker-Planck equation.

Finally, one can also compute the propagator $G_{>}^{\rmd}(x,0,n)$ associated to a RW which is constrained to stay strictly above $0$. To do so, one can simply analyze the first step. This first step is necessarily a step $+1$, such that $\eta(1) = +1$, which happens with probability $1/2$. After this first step, the remaining propagator is just given by $G_{\geq}^\rmd(x-1,0,n-1)$. Therefore, one has
\beq\label{eq:G>d}
&&G_{>}(x,0,0) = \delta_{x,0} \\
&&G_{>}(x,0,n) = \frac{1}{2} G_{\geq}^\rmd(x-1,0,n-1) \;, \; n \geq 1 \;.
\eeq
One thus obtains the generating function $\tilde G_{>}(x,0,z)$ as
\beq\label{GF_G>d}
\tilde G_{>}(x,0,z) = \sum_{n=0}^\infty G_{>}(x,0,n) z^n = \left(\frac{1-\sqrt{1-z^2}}{z} \right)^{x} \;.
\eeq

\section{Useful formulas for RWs with exponential jump distribution}\label{sec:appendix_exp}

Here we consider a RW, starting from $x(0) = 0$ and evolving via
\beq\label{def_exp_app}
x(k) = x(k-1) + \eta(k) \;,
\eeq
where the jump variables $\eta(k)$'s are i.i.d. random variables distributed according to $p^{\e}(\eta) = \frac{1}{2\,b} \e^{-|\eta|/b}$. A remarkable feature of the exponential jump distribution is that the constrained Green's function $G_>^{\rm e}(x,x_0,n)$ can be computed exactly \cite{CM05}. 

To begin with, it is useful to compute the Fourier transform of the jump distribution
\beq\label{eq:Fourier}
\hat p^{\e}(q) = \int_{-\infty}^\infty \rmd \eta \, p^\e(\eta)\, \e^{i\, q\,\eta} = \frac{1}{1 + (b \, q)^2} \;,
\eeq
from which we easily obtain the free propagator 
\beq\label{eq:Ge_app}
G^{\e}(x,x_0,n) = \int_{-\infty}^\infty \frac{\rmd q}{2 \pi} \, \e^{-i \, q (x-x_0)} \frac{1}{\left[1 + (b\,q)^2 \right]^n}.
\eeq
Finally, the GF of $G^{\e}(x,x_0,n)$ is given by
\beq\label{eq:Ge_app2}
\hspace*{-1.5cm}\tilde G^{\e}(x,x_0,z) = \sum_{n=1}^\infty G^{\e}(x,x_0,n) z^n = z \, \int_{-\infty}^\infty \frac{\rmd q}{2 \pi} \, \e^{-i \, q (x-x_0)} \frac{1}{1-z + (b\,q^2)} \;,
\eeq
where the Dirac delta function $\delta(x)$ comes from the term $n=0$, while the second one comes from the sum from $n=1$ to $\infty$ of the geometric series. Finally, the integral over $q$ in (\ref{eq:Ge_app2}) can be explicitly evaluated to yield the result given in Eq. (\ref{eq:expr_GG>_exp})
\beq\label{eq:Ge_app3}
\tilde G^\e(x,0,z) = \sum_{n=1}^\infty z^n G^\e(x,0,z) = \frac{z}{2b\sqrt{1-z}} \e^{-\frac{|x|}{b}\sqrt{1-z}} \;.
\eeq

We now come to the constrained propagator $G_>^\e(x,0,n)$, which is harder to compute. The easiest way to compute it is to use the so-called Hopf-Ivanov formula \cite{Ivanov} (see also \cite{MMS2014} for a detailed derivation of this result) which gives the following expression
\beq\label{Ivanov}
\sum_{n=0}^\infty z^n \int_0^\infty \rmd x \, G_>^\e(x,0,n) \, \e^{-\lambda \, x} = \phi(\lambda,z) \;,
\eeq
where $\phi(\lambda,z)$ is given by
\beq\label{expr_phi}
\phi(\lambda,z) = \exp{\left(-\frac{\lambda}{\pi} \int_0^\infty \rmd q \frac{\ln\left(1- z \, \hat p^{\e}(q) \right)}{\lambda^2+q^2} \right)} \;,
\eeq 
where $\hat p^\e(q)$ is the Fourier transform of the jump distribution given in (\ref{eq:Fourier}). The function $\phi(\lambda,z)$ is thus given by
\beq\label{expr_phi2}
\phi(\lambda,z) = \exp{\left(-\frac{\lambda}{\pi} \int_0^\infty \rmd q \frac{\ln\left(\frac{1-z+(b\,q)^2}{1+(b\,q)^2} \right)}{\lambda^2+q^2} \right)} \;.
\eeq 
The integral over $q$ in Eq. (\ref{expr_phi2}) can be explicitly evaluated, using the formula 4.295-7 of Ref. \cite{Grad}
\beq\label{eq:integral_log}
\int_0^\infty \rmd x \, \frac{\ln{(a^2+b^2x^2)}}{c^2+d^2 x^2} = \frac{\pi}{c \,d} \ln{\left(\frac{a\,d+b\,c}{d}\right)} \;,\; {\rm for}\;\; a,b,c,d>0 \;,
\eeq
to yield finally the identity
\beq\label{eq:ivanov_explicit}
\hspace*{-1.cm}\sum_{n=0}^\infty z^n \int_0^\infty \rmd x \, G_>^\e(x,0,n) \, \e^{-\lambda \, x} = \frac{1+\lambda \, b}{\sqrt{1-z}+\lambda\,b} = 1+\frac{1-\sqrt{1-z}}{\sqrt{1-z}+\lambda\,b} \;.
\eeq
It is then easy to inverse the Laplace transform with respect to $x$ to obtain $\tilde G_>^\e(x,0,z)$ given in Eq. (\ref{eq:expr_GG>_exp2})
\beq
\tilde G^\e_>(x,0,z) = \sum_{n=0}^\infty z^n G^\e_>(x,0,z) = \delta(x) + \frac{1-\sqrt{1-z}}{b} \e^{-\frac{|x|}{b}\sqrt{1-z}} \;. \label{eq:expr_GG>_app}
\eeq
Note finally that, by differentiating the relation in Eq. (\ref{eq:ivanov_explicit}) $(m-2)$ times with respect to $\lambda$ and setting $\lambda = 1/b$, one obtains straightforwardly the relation given in Eq. (\ref{eq:GFq}).

\section{Analysis of the distribution of $\ell_{\max,B}^{\rm e}(n)/n$}\label{sec:appendix_PDF}

In this appendix, we derive the behavior for the PDF $f^\e_{\cal R}(r)$ first for $r \in [1/2,1]$ and then for $r \to 0$ given in the text in Eqs. (\ref{eq:explicitfR}) and (\ref{eq:essential}) respectively. 

\subsection{Density $f^\e_{\cal R}(r)$ in the interval $[1/2,1]$}

We start from Eq.~(\ref{eq:scalingFe}) given in the text
\beq\label{eq:app_f_1}
\fl \sum_{n\ge0}\e^{-s \,n} F^\e(\l,n)_{\B}
=\frac{1}{b\, \sqrt{s}} I(s\l) \;,
\eeq
valid in the limit $s \to 0$, $\ell \to \infty$, keeping the product $x = s \, \ell$ fixed and where the function $I(t)$ is given in Eq. (\ref{eq:def_I}). In the limit $s \to 0$, the discrete sum in Eq.~(\ref{eq:app_f_1}) can be replaced by an integral and $F^\e(\l,n)_{\B}$ can be written as
\beq
F^\e(\l,n)_{\B}=\int_{\Gamma}\frac{\d s}{2 {\rm i} \pi}\e^{n s}\frac{1}{b\sqrt{s}}I(s\l) \;,
\eeq
where $\Gamma$ is a Bromwich contour. Note that by definition, we have
\beq
F^\e(\l,n)_{\B}=\prob(\l^\e_{\max,B}(n)<\l)_{\B}=\prob\Big(\frac{n}{\l^\e_{\max,B}}>\frac{n}{\l}\Big)_{\B} \;.
\eeq
In the limit of large $n$ this probability converges to the complementary distribution function of the random variable ${\cal V}= 1/{\cal R} = \lim_{n\to\infty}n/\l_{\max}(n)$,
\beq
\bar F^\e_{\cal V}(v)_{\B}=\prob\Big({\cal V}>v\Big)_{\B}=
\frac{\sqrt{v}}{b\sqrt{n}}
\int\frac{\d x}{2{\rm i} \pi}\e^{x v}\frac{1}{\sqrt{x}}I(x),
\eeq
with $v=n/\l$  and $x=s\,\l$ as above.
Finally, dividing by $G^\e(0,0,n) \sim 1/(2 b \sqrt{\pi n})$,
\beq\label{eq:barFV}
\bar F^\e_{\cal V}(v)=
2\sqrt{\pi v}
\int\frac{\d x}{2{\rm i} \pi}\e^{x v}\frac{1}{\sqrt{x}}I(x).
\eeq
We want to analyze the behavior of $F^\e_{\cal R}(r)$ for $r$ close to $1$, or equivalently the small $v$ behavior of 
of $\bar F^\e_V(v)$, which is dominated by the large $x$ behavior of the integrand in the Bromwich integral in Eq. (\ref{eq:barFV}).
Thus we need to expand $I(x)$ given by (\ref{eq:def_I}) in the small parameter 
$\epsilon(x)=\e^{-x}$.
Formally
\beq
\fl\int\frac{\d x}{2{\rm i} \pi}\e^{x v}\frac{1}{\sqrt{x}}I(x)
\approx 
\int\frac{\d x}{2{\rm i} \pi}\e^{x v}\frac{1}{\sqrt{x}}\left({\cal I}_0(x)+ {\cal I}_1(x)\,\e^{-x}+ {\cal I}_2(x)\,\e^{-2x}+\dots\right)\;,
\eeq
where the functions ${\cal I}_k(x)$ admits an expansion in powers (both positive and negative) of $\sqrt{x}$. Therefore the first term $\propto \e^{-x}$ contributes to the function for $v>1$, the second term, $\propto \e^{-2\,x}$, for $v>2$, and so on. At first order in $\epsilon(x)=\e^{-x}$ we have
\beq\label{eq:Iapprox}
I(x)\approx\frac{1}{2}+\frac{1}{4}
\left((4x-1)\frac{\e^{-x}}{\sqrt{\pi x}}-(4x+1)\erfc \sqrt{x}\right) + {\cal O}(\e^{-2 x}) \;,
\eeq
which, divided by $\sqrt{x}$, can be Laplace inverted to give for $\bar F_V(v)$
\beq
\bar F^\e_{\cal V}(v)=\frac{1}{2}+\frac{1}{v}-\frac{\sqrt{v}}{2} \qquad (1\le v\le 2) \;,
\eeq
and therefore
\beq
f^\e_{\cal V}(v)= - \frac{\rmd}{\rmd v} \bar F^\e_V(v) = \frac{1}{v^2}+\frac{1}{4\sqrt{v}}\qquad (1\le v\le 2) \;.
\eeq
Finally, using $f^\e_{\cal R}(r) = f^\e_{\cal V}(1/r)/r^2$, one obtains
\beq\label{eq:prediction}
f^\e_{\cal R}(r)=1+\frac{1}{4 r^{3/2}}\qquad \Big(\frac{1}{2}\le r\le 1\Big) \;,
\eeq
as announced in the text in Eq. (\ref{eq:explicitfR}). 

\subsection{Small $r$ behavior of $f^\e_{\cal R}(r)$}

We now study the small $r$ behavior of $f^\e_{\cal R}(r)$, or equivalently the large $v$ behavior of $f^\e_{\cal V}(v)$ (recalling
that ${\cal V} = 1/{\cal R}$). One can show that the function $I(x)/\sqrt{x}$ is an entire in the complex $x$ plane.
Hence the integral~(\ref{eq:barFV}) is expected to be dominated by a saddle point in $x$.
Let us define 
\beq
\phi(x)=\ln \frac{I(x)}{\sqrt{x}}.
\eeq
The saddle-point equation is $v+\phi'(x)=0$, 
thus $\phi'(x)$ is large and negative, hence $x$ is large and negative too.
Setting $x=-y^2$ we obtain the estimate the large negative $x$ behavior of $F(x)$ in Eq. (\ref{eq:scaling_h_discrete}) as
\beq
F(x=-y^2)\approx{\rm i}\frac{\e^{y^2}}{2\sqrt{\pi}y^3},
\eeq
which yields
\beq
\phi(x=-y^2)\sim \frac{\e^{2y^2}}{4\pi y^4}=\frac{\e^{-2x}}{4\pi x^2},
\eeq
coming from the dominant term $\e^{x(F(x))^2}$ in $I(x)$.
The saddle-point equation now gives
\beq
v\sim \e^{-2x},
\eeq
hence, for $v$ large, using estimates at exponential order, we find
\beq\label{eq:vlarge}
 f^\e_{\cal V}(v) = -\frac{\rm d}{\rm d v}\bar F^\e_{\cal V}(v)\sim \e^{-\frac{1}{2}v\ln v},
\eeq
i.e., a super-exponential decay for this quantity.
As a consequence one can also predict that the essential singularity at the origin of $f^\e_{\cal R}(r) = f^\e_{\cal V}(1/r)/r^2$ is given by
\beq
f_R(r)\sim \e^{\ln r/(2 r)} \;,
\eeq
as given in the text in Eq. (\ref{eq:explicitfR}).

%\newpage
\section*{References}
%\vspace*{0.5cm}

\end{document}